\documentclass[useAMS,usenatbib]{mn2e}
\usepackage{graphicx,color,afterpage}
\usepackage[total={17.8cm,24.0cm},centering]{geometry}
\usepackage{times,subfigure,rotating}

\def\hi{\mbox{H\,\sc{i}}}

\def\atlas{{{ATLAS}}$^{\rm 3D}$}
\def\mum{$\mu$m}
\def\kms{km s$^{-1}$}

\def\msun{M$_{\odot}$}

\def\arcsec{$^{\prime \prime}$}
\definecolor{Mygrey}{gray}{0.75}

\newcommand{\gtsimeq}{\raisebox{-0.6ex}{$\,\stackrel{\raisebox{-.2ex}{$\textstyle >$}}{\sim}\,$}}
\newcommand{\farc}{\mbox{\ensuremath{.\!\!^{\prime\prime}}}}

\usepackage[compact]{titlesec}
\titlespacing{\section}{0pt}{*2}{*1}
\title[Molecular gas in dust lane ETGs]{Molecular and atomic gas in dust lane early-type galaxies - I: Low star-formation efficiencies in minor merger remnants} 
\author[Timothy A. Davis et al.]{\parbox{\textwidth}{Timothy~A.~Davis$^{1,2}$\thanks{E-mail: \texttt{t.davis4@herts.ac.uk}}, Kate Rowlands$^{3}$, James R. Allison$^{4}$, Stanislav S. Shabala$^{5}$, Yuan\mbox{-}Sen Ting$^{6}$, Claudia  del P. Lagos$^{1,7}$, Sugata Kaviraj$^{2}$,
Nathan Bourne$^{8}$,
Loretta Dunne$^{8,9}$,
Steve Eales$^{10}$,
Rob.\,J.~Ivison$^{1,8}$, Steve Maddox$^{8,9}$,
Daniel~J.~B.~Smith$^{2}$, Matthew.~W.~L. Smith$^{10}$, and Pasquale Temi$^{11}$}
\vspace{0.4cm}\\
\parbox{\textwidth}{$^{1}$European Southern Observatory, Karl-Schwarzschild-Str. 2, 85748 Garching, Germany\\
$^{2}$Centre for Astrophysics Research, University of Hertfordshire, Hatfield, Herts AL1 9AB, UK\\
$^{3}$School of Physics and Astronomy, University of St Andrews, North Haugh, St Andrews KY16 9SS, UK\\
$^{4}$CSIRO Astronomy \& Space Science, P.O. Box 76, Epping NSW 1710, Australia\\
$^{5}$School of Mathematics and Physics, University of Tasmania, Private Bag 37, Hobart TAS 7001, Australia\\
$^{6}$Harvard-Smithsonian Center for Astrophysics, 60 Garden Street, Cambridge, MA 02138, USA\\
$^{7}$International Centre for Radio Astronomy (ICRAR), University of Western Australia, Crawley, WA6009, Australia \\ 
$^{8}$Institute for Astronomy, The University of Edinburgh, Royal Observatory, Blackford Hill, Edinburgh EH9 3HJ, UK \\
$^{9}$Department of Physics and Astronomy, University of Canterbury, Private Bag 4800, Christchurch, New Zealand \\ 
$^{10}$School of Physics \&\ Astronomy, Cardiff University, Queens Buildings, The Parade, Cardiff, CF24 3AA, UK\\
$^{11}$Astrophysics Branch, NASA/Ames Research Center, MS 245-6, Moffett Field, CA 94035, USA }}
\begin{document}

\date{Accepted 2015 March 17. Received 2015 March 12; in original form 2015 January 12}

\pagerange{\pageref{firstpage}--\pageref{lastpage}} \pubyear{2012}

\maketitle

\label{firstpage}

\begin{abstract}

In this work we present IRAM-30m telescope observations of a sample of bulge-dominated galaxies with large dust lanes, which have had a recent minor merger.
We find these galaxies are very gas rich, with H$_2$ masses between 4$\times$10$^{8}$ and 2$\times$10$^{10}$ \msun. We use these molecular gas masses, combined with atomic gas masses from an accompanying paper, to calculate gas-to-dust and gas-to-stellar mass ratios.  The gas-to-dust ratios of our sample objects vary widely (between $\approx$50 and 750), suggesting many objects have low gas-phase metallicities, and thus that the gas has been accreted through a recent merger with a lower mass companion. We calculate the implied minor companion masses and gas fractions, finding a median predicted stellar mass ratio of $\approx$40:1. The minor companion likely had masses between $\approx$10$^7$ - 10$^{10}$ \msun. The implied merger mass ratios are consistent with the expectation for low redshift gas-rich mergers from simulations.
We then go on to present evidence that (no matter which star-formation rate indicator is used) our sample objects have very low star-formation efficiencies (star-formation rate per unit gas mass), lower even than the early-type galaxies from \atlas\ which already show a suppression. This suggests that minor mergers can actually suppress star-formation activity. We discuss mechanisms that could cause such a suppression, include dynamical effects induced by the minor merger.

\end{abstract}


\section{Introduction}

Early type galaxies (ETGs) have traditionally been thought of as gas-poor objects. This perception has slowly been challenged (e.g. \citealt{1991ApJS...75..751R}), and recent results (e.g. \citealt{2007MNRAS.377.1795C,Welch:2010in,2011MNRAS.414..940Y}) have showed with statistically complete samples that $\approx$1/4 of ETGs have $>$10$^7$ \msun\ of cold molecular gas, and $\approx$40\% of ETGs have sizeable atomic gas reservoirs \citep{2006MNRAS.371..157M,Sage:2006ko,2012MNRAS.422.1835S}. Dust is also present in a large fraction of ETGs, e.g. \cite{2012ApJ...748..123S} find that $\approx$50\% of ETGs in the \emph{Herschel} Reference Survey have cold dust. 
Low level residual star formation has also been detected through studies of UV emission \citep[e.g.][]{2005ApJ...619L.111Y,2007ApJS..173..619K,2010ApJ...714L.290S,Wei:2010bt}, optical emission lines \citep[e.g.][]{Crocker:2011ic} and infrared emission (e.g. \citealt{1989ApJS...70..329K,2007MNRAS.377.1795C,2009ApJ...695....1T, 2010MNRAS.402.2140S,2014MNRAS.444.3427D}).

Observationally, in field environments, the gas and dust in ETGs seems to come mainly from external sources.
The evidence for this comes from the kinematic decoupling of the cold and ionized gas from the stars \citep{Sarzi:2006p1474,2011MNRAS.417..882D,2014MNRAS.438.2798K} , the orders of magnitude surplus of ISM compared to what one would expect from stellar mass loss (e.g. \citealt{1998A&A...338..807M,1989ApJS...70..329K,2012MNRAS.419.2545R,2012MNRAS.423...49K}), the large spread in gas-to-dust ratios within their cold ISM \citep{2012ApJ...748..123S}, the large fraction of gas rich ETGs that are morphologically disturbed in deeper imaging (e.g. \citealt{2005AJ....130.2647V,2015MNRAS.446..120D}) and the presence of young kinematically decoupled cores in IFU surveys (e.g. \citealt{McDermid:2006bj}).

The external sources that supply this gas likely include both mergers with gas-rich objects and accretion of gas from the intergalactic medium.
Major mergers are thought to be too rare in the low redshift universe to supply enough gas, so minor mergers are argued to be the dominant source of gas \citep[e.g.][]{2009MNRAS.394.1713K,2011MNRAS.411.2148K}. Indeed, some studies claim that minor mergers are likely to be driving at least half of the cosmic star formation budget in the local Universe \citep{2014MNRAS.440.2944K,2014MNRAS.437L..41K}. We thus need a better understanding of the role of minor mergers, in ETGs in particular, and galaxy evolution in general.

The nearest large ETG to us, Centaurus A, is a good example of this process in action. Its 10$^9$ \msun\ of cold gas and 10$^7$ \msun\ of dust were likely acquired in a recent, relatively minor merger \citep{1998A&ARv...8..237I,2012MNRAS.422.2291P}. This gas/dust reservoir makes its presence obvious in optical images as a large warped dust lane, seen in absorption against the stellar continuum. By selecting disturbed ETGs in the field with such large dust lanes we are able to probe the minor merging process, which supplies the fuel that is rejuvenating these objects.  

Theoretically it has been argued that the main mode of gas supply in ETGs in the local universe is accretion of gas from the inter-galactic and circum-galactic medium \citep{2014MNRAS.443.1002L}, with minor mergers playing a secondary, albeit still very relevant role. Understanding if the observational and theoretical models for gas accretion onto massive galaxies can be reconciled is an important area of future research \citep[e.g.][]{2015MNRAS.448.1271L}.

Dust-lane ETGs are relatively rare and the properties of the gas in such objects have yet to be studied in a statistical sample. In this work we present molecular gas observations of a sample of bulge-dominated galaxies with large far infrared bright dust lanes, selected from the unbiased Sloan Digital Sky Survey (SDSS) photometry in \cite{2012MNRAS.423...49K} as part of the Galaxy Zoo project \citep{2008MNRAS.389.1179L}. See Section \ref{sample} for the full selection criteria. These objects are all recent minor merger candidates \citep{2002AJ....123..729O,2012MNRAS.423...59S}. We selected the most dust rich objects (using 250\mum\ luminosities) and observed them with the Institut de Radioastronomie MillimŽt\'erique (IRAM) 30m telescope, in order to estimate the mass of molecular gas present in each system. 

In this paper we present the IRAM-30m observations (see Section \ref{data}), and the derived quantities (see Section \ref{results}). In Section \ref{discuss} we discuss the dust-to-gas and dust-to-stellar mass ratios of these systems, derived from \emph{Herschel}\footnote{{\it Herschel} is an ESA space observatory with science instruments provided by European-led Principal Investigator consortia and with important participation from NASA.} space telescope \citep{2010A&A...518L...1P} observations from the \emph{Herschel}-ATLAS survey \citep[H-ATLAS;][]{2010PASP..122..499E}. We use these quantities to constrain the mass of the dwarf companion systems (and thus the mass ratio of the mergers). We also explore how the minor merging process affects the efficiency with which gas turns into stars. Finally we conclude in Section \ref{conclude}. 
In this paper we have assumed a cosmology with H$_0$= 71 km s$^{-1}$ Mpc$^{-1}$, $\Omega_m$= 0.27 and $\Omega_\Lambda$=0.73.

\section{Sample}
\label{sample}

\begin{table*}
\caption{Properties of the galaxies observed in this work}
\begin{tabular*}{0.9\textwidth}{@{\extracolsep{\fill}}l r r r r r r}
\hline
NED name & DETG ID & $z$ & D$_{\rm lum}$ & M$_{\rm *}$ & M$_{\rm d}$ & SFR$_{\rm MAGPHYS}$  \\
& & & Mpc & log$_{10}$(M$_{\odot}$) & log$_{10}$(M$_{\odot}$) & (\msun\ yr$^{-1}$) \\
 (1) & (2) & (3) & (4) & (5) & (6) & (7)\\
 \hline
2MASXJ09033081-0106127 &         18 &      0.040 &      174.2 &       10.9$\pm$       0.1 &       7.41$\pm$      0.05 &       0.09$^{+0.05}_{-0.01}$\\
2MASXJ14161186+0152048 &        106 &      0.082 &      368.3 &       11.2$\pm$       0.1 &       8.11$\pm$      0.03 &       0.02$^{+0.06}_{-0.01}$\\
CGCG018-061 &        231 &      0.030 &      129.7 &       10.5$\pm$       0.2 &       7.65$\pm$      0.08 &       0.65$^{+0.15}_{-0.17}$\\
2MASXJ13341710+3455455 &        920 &      0.025 &      107.6 &       10.8$\pm$       0.1 &       7.28$\pm$      0.07 &       0.05$^{+0.03}_{-0.03}$\\
NGC5233 &        967 &      0.026 &      112.0 &       11.1$\pm$       0.1 &       7.59$\pm$      0.08 &       0.02$^{+0.04}_{-0.01}$\\
2MFGC10709 &        984 &      0.037 &      160.8 &       10.8$\pm$       0.1 &       7.43$\pm$      0.07 &       0.08$^{+0.03}_{-0.04}$\\
UGC08505 &       1004 &      0.040 &      174.2 &       10.8$\pm$       0.1 &       8.12$\pm$      0.06 &       0.87$^{+0.40}_{-0.50}$\\
UGC08536 &       1014 &      0.025 &      107.6 &       10.6$\pm$       0.1 &       7.43$\pm$      0.06 &       1.28$^{+0.41}_{-0.43}$\\
2MASX J13194502+3134277 &       1030 &      0.045 &      196.7 &       10.7$\pm$       0.1 &       7.41$\pm$      0.03 &     0.01$^{+0.01}_{-0.01}$\\
2MASXJ13411704+2616190 &       1111 &      0.064 &      283.7 &       11.1$\pm$       0.1 &       8.18$\pm$      0.07 &       0.68$^{+0.21}_{-0.29}$\\
NGC4797 &       1232 &      0.026 &      112.0 &       11.2$\pm$       0.1 &       7.00$\pm$      0.06 &       0.03$^{+0.03}_{-0.01}$\\
2MASXJ12581769+2708117 &       1265 &      0.040 &      174.2 &       10.8$\pm$       0.1 &       7.45$\pm$      0.08 &       0.09$^{+0.03}_{-0.02}$\\
2MASXJ13010083+2701312 &       1267 &      0.078 &      349.3 &       10.9$\pm$       0.1 &       7.95$\pm$      0.08 &       0.44$^{+0.44}_{-0.37}$\\
2MASXJ13333299+2616190 &       1268 &      0.037 &      160.8 &       10.8$\pm$       0.1 &       7.04$\pm$      0.03 &       0.01$^{+0.06}_{-0.01}$\\
2MASX J12561089+2332394 &       1362 &      0.133 &      618.6 &       11.0$\pm$       0.1 &       8.24$\pm$      0.03 &       2.51$^{+1.48}_{-1.50}$\\
2MASXJ13244786+2417386 &       1372 &      0.065 &      288.4 &       10.7$\pm$       0.1 &       7.80$\pm$      0.08 &       0.23$^{+0.25}_{-0.16}$\\
CGCG013-092 &       1947 &      0.035 &      151.9 &       10.9$\pm$       0.1 &       7.79$\pm$      0.06 &       0.28$^{+0.06}_{-0.04}$\\
   \hline
\end{tabular*}
\parbox[t]{0.9 \textwidth}{ \textit{Notes:}  Column 1 shows the NED designator for these objects. Column 2 shows the DETG ID number for these sources from K13. Column 3 and 4 show the spectroscopic redshift, and associated luminosity distance (assuming H$_0$= 71 km s$^{-1}$ Mpc$^{-1}$, $\Omega_m$= 0.27 and $\Omega_\Lambda$=0.73). The stellar mass, dust mass and star-formation rate (averaged over the last 10$^8$ years) of these objects are in Column 5, 6 and 7, derived as in K13.}
\label{proptable}
\end{table*}

The \citet[][hereafter K12]{2012MNRAS.423...49K} dust lane sample is drawn from the entire SDSS Data Release 7 (which
covers 11663 deg$^{2}$) through visual inspection of galaxy images
via the Galaxy Zoo \citep[GZ;][]{2008MNRAS.389.1179L} project. Approximately 19,000 GZ galaxies at $z < 0.1$ were
flagged as containing a dust lane by \emph{at least} one GZ user
(individual galaxies have 50+ classifications), and each galaxy was re-inspected by our team, as described in K12.
This yields a final sample of 352 dust lane galaxies, with a median redshift, absolute $r$-band magnitude and stellar
mass of 0.04, -21.5 and 6$\times$10$^{10}$ M$_{\odot}$ respectively. 
These objects are massive, have red optical colours and are bulge-dominated. Due to the Galaxy Zoo selection criteria, classical ellipticals, lenticular galaxies, and some massive bulge-dominated Sa-type spirals are included. Despite this, we refer to these objects as dust lane early-type galaxies (DETGs) in this work to be consistent with previous studies.

The DETGs in this study were extracted from this parent sample by cross-matching the K12
objects with the Phase 1 and North Galactic Pole (NGP) fields of
the \emph{Herschel}-ATLAS survey \citep[H-ATLAS;][]{2010PASP..122..499E}, as described in \citet[][hereafter K13]{2013MNRAS.435.1463K}. 
The final DETG sample of K13 comprises 23 galaxies,
with a median redshift of 0.05 and stellar mass of 8$\times$10$^{10}$
M$_{\odot}$ respectively. The sample of DETGs discussed in this paper are the 17 dustiest objects (estimated using the 250-500$\mu$m data) from the K13 sample. 
These objects are all recent minor merger candidates, having clear tidal low surface brightness features visible in their optical images (see \citealt{2012MNRAS.423...59S} for full details).

 \subsection{Galaxy physical parameters}
 \label{sfrcomp}

Twenty filter photometry is available for each of these objects, covering three
orders of magnitude in wavelength (0.12 - 500 $\mu$m; \citealt{2010MNRAS.409...38I,2011MNRAS.415..911P,2011MNRAS.415.2336R,2011MNRAS.416..857S, 2011MNRAS.413..971D}). These data are used in K13 to estimate the
 physical properties of our DETGs using the
energy-balance technique of \citet[][DC08 hereafter]{2008MNRAS.388.1595D} via the \texttt{MAGPHYS} code\footnote{The DC08 models are publicly available as a user-friendly model package {\sc magphys} at www.iap.fr/magphys/.}. 
{ We refer the reader to K13 for more details, but outline below the basic premise of these models.}

{ In each model it is assumed that the UV--optical radiation emitted by stellar populations is partially absorbed by dust, and this absorbed energy is matched to that re-radiated in the FIR. The optical library of 50,000 spectra is produced using the the latest version of the population synthesis code of \citet{2003MNRAS.344.1000B}, Charlot \& Bruzual (2007, in prep), and assumes exponentially declining SFHs with superimposed stochastic bursts. The model spectra cover a wide range of age, metallicity, SFH and dust attenuation and assume a \citet{2003PASP..115..763C} IMF. The infrared libraries contain 50,000 SEDs comprised of four different temperature dust components, from which the dust mass is calculated. In stellar birth clouds, these components are polycyclic aromatic hydrocarbons (PAHs), hot dust (stochastically heated small grains with a temperature $130-250$\,K), and warm dust in thermal equilibrium ($30-60$\,K). In the diffuse ISM the relative fractions of these three dust components are fixed, but an additional cold dust component with an adjustable temperature between 15 and 25\,K is added. The dust mass absorption coefficient $\kappa_{\lambda} \propto \lambda^{-\beta}$ has a normalisation of $\kappa_{850}=0.077\,\rm{m}^2\,\rm{kg}^{-1}$. A dust emissivity index of $\beta=1.5$ is assumed for warm dust, and $\beta=2.0$ for the cold dust component.}

The physical properties of our sample DETGs are presented and discussed in detail in K13, and here we use similar values derived using the same modelling code, but with updated PACS and SPIRE photometry (Valiante et al., in prep). The galaxy properties (and common names extracted from the NASA/IPAC Extragalactic Database; NED\footnote{http://ned.ipac.caltech.edu/}) are presented in Table \ref{proptable}. We note that the changes caused by using the updated photometry are generally small, and the results discussed here are robust to the inclusion of this reprocessed \emph{Herschel} data.

\section{IRAM-30m Data}
\label{data}

The IRAM 30-m telescope at Pico Veleta, Spain, was used between the 4th and 7th of January 2013 to observe CO emission in our sample of dust rich DETGs (proposal 182-12, PI Davis/Kaviraj). 
 We aimed to detect CO(1-0) and CO(2-1) simultaneously in the 3mm and 1mm atmospheric windows.
The beam full width at half-maximum (FWHM) of the IRAM-30m at the frequency of these lines is 21\farc3 and 10\farc7, respectively. The Eight MIxer Receiver (EMIR) was used for observations in the wobbler switching mode, with reference positions offset by $\pm$100\arcsec\ in azimuth. The Fourier Transform Spectrograph (FTS) back-end gave an effective total bandwidth of $\approx$4 GHz per window, and a raw spectral resolution of 200 kHz ($\approx$0.6 \kms\ at $\lambda=$3mm, $\approx$0.3 \kms\  at $\lambda=$1mm). The Wideband Line Multiple Autocorrelator (WILMA) back-end was used simultaneously with the FTS, and WILMA data were used at 230 GHz when possible to avoid known issues with platforming in the FTS.

The system temperatures ranged between 90 and 200 K at 3~mm and between 100 and 220 K at 1~mm.
 The pointing was checked every 2 hours on a nearby planet or bright quasar, and the focus was checked at the beginning of each night, after sunrise, or more often if a suitable planet was available. The time on source ranged from 15 to 190 min, being weather-dependent, and was interactively adjusted to try and ensure detection of molecular emission.
 
The individual $\approx$6 minute scans were inspected, and baselined, using a zeroth-, first- or second-order polynomial, depending on the scan. Scans with poor quality baselines or other problems were discarded.
The good scans were averaged together, weighted by the inverse square of the system temperature. We consider emission lines where the integrated intensity has greater than a 3$\sigma$ significance (including the baseline uncertainty;  \citealt{2012MNRAS.421.1298C}) to be detected.  

Integrated intensities for each detected line in each galaxy were computed by fitting a Gaussian profile in the \texttt{\small CLASS} package of \texttt{\small GILDAS}\footnote{http://www.iram.fr/IRAMFR/GILDAS/ - accessed 14/01/13}, or summing the spectrum over the velocity (where the profile shape is non-Gaussian). Both methods produce consistent results, and here we present the values derived from the sum. 
We convert the spectra from the observed antenna temperature (T$_a^*$) into units of Jy, utilising the point source conversion S$_{\rm Jy/K}$ as tabulated on the IRAM website\footnote{http://www.iram.es/IRAMES/mainWiki/EmirforAstronomers - accessed 23/04/2013}. Where the exact frequency has no measured efficiency available we linearly interpolated the values required.
Table \ref{obstable} lists the RMS noise levels in each spectrum, the velocity width summed over, and the line integrated intensities (in Jy \kms). Converting these integrated line intensities to the main beam temperature scale can be accomplished by dividing by a factor of 1.28 at 3mm\S.

\begin{figure*}
\begin{minipage}[t]{1.0\textwidth}
\begin{center}
$\begin{array}{cccc}
\begin{turn}{90}\large \hspace{1.1cm} DETG0018\end{turn} &
\includegraphics[height=3.9cm,angle=0,clip,trim=0.0cm -0.0cm 0cm 0cm]{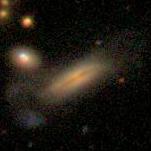}  &
\includegraphics[height=4cm,angle=0,clip,trim=0.0cm 0cm 0cm -0.3cm]{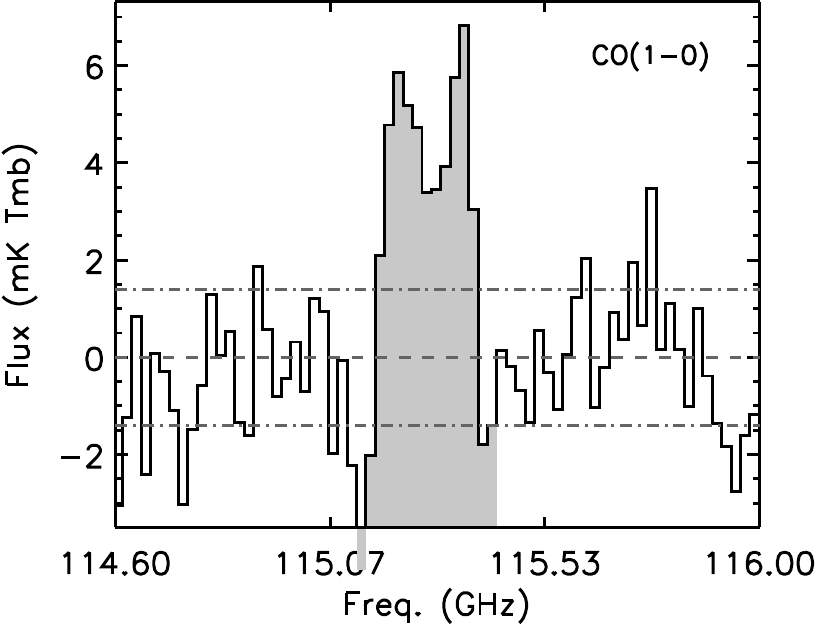} &
\includegraphics[height=4cm,angle=0,clip,trim=0.0cm 0.0cm 0cm -0.3cm]{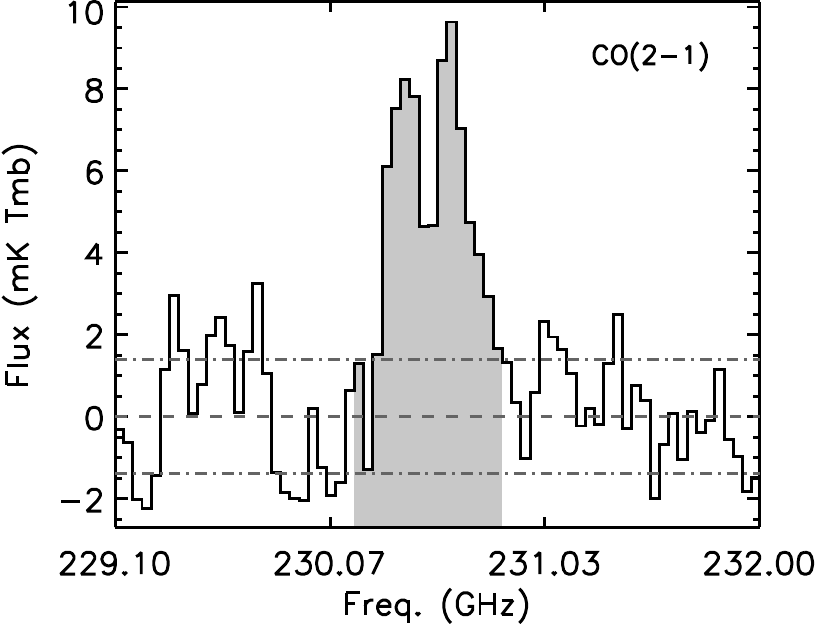} \\
\begin{turn}{90}\large \hspace{1.1cm} DETG0106\end{turn} &
\includegraphics[height=3.9cm,angle=0,clip,trim=0.0cm 0.0cm 0cm 0cm]{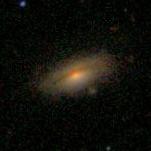} &
\includegraphics[height=4cm,angle=0,clip,trim=0.0cm 0.0cm 0cm -0.3cm]{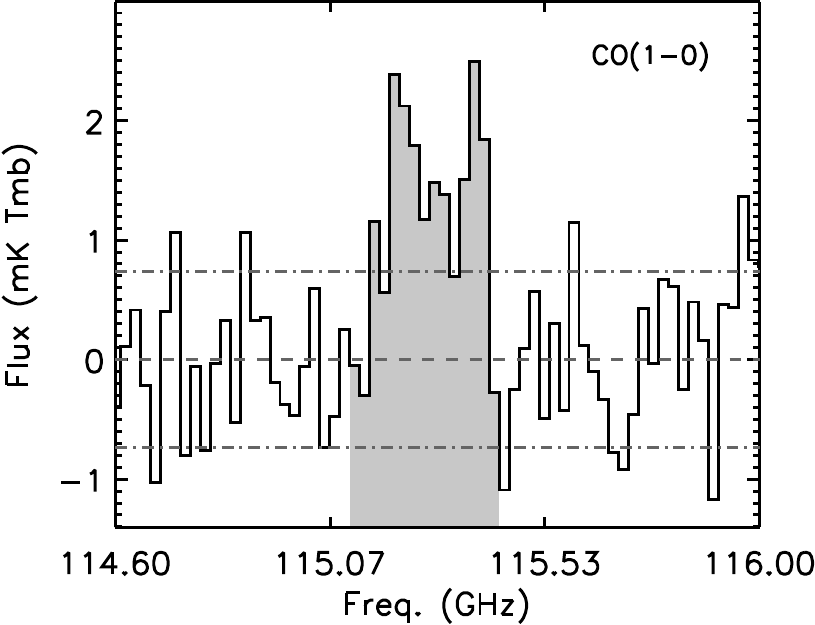} &
\includegraphics[height=4cm,angle=0,clip,trim=0.0cm 0.0cm 0cm -0.3cm]{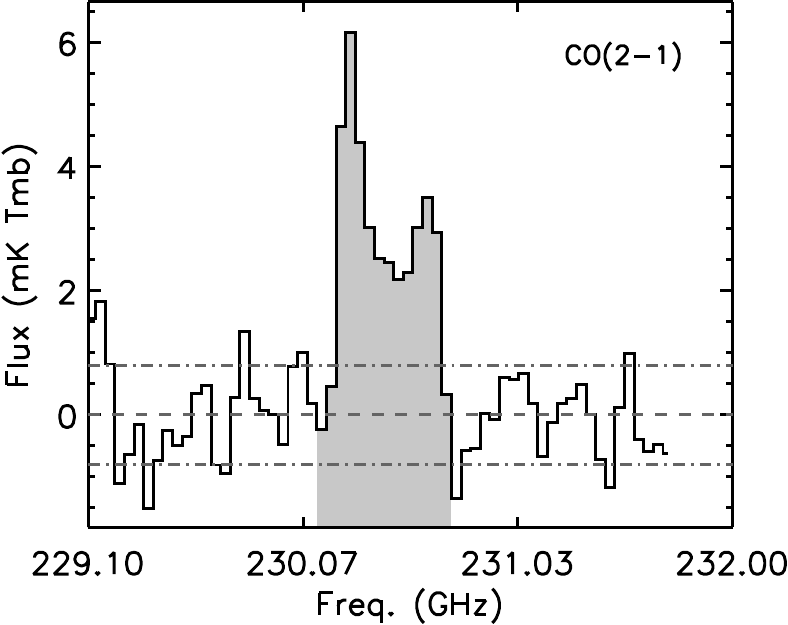} \\
\begin{turn}{90}\large \hspace{1.1cm} DETG0231 \end{turn} &
\includegraphics[height=4cm,angle=0,clip,trim=0.0cm 0.0cm 0cm 0cm]{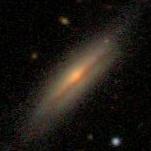} &
\includegraphics[height=4cm,angle=0,clip,trim=0.0cm 0.0cm 0cm -0.3cm]{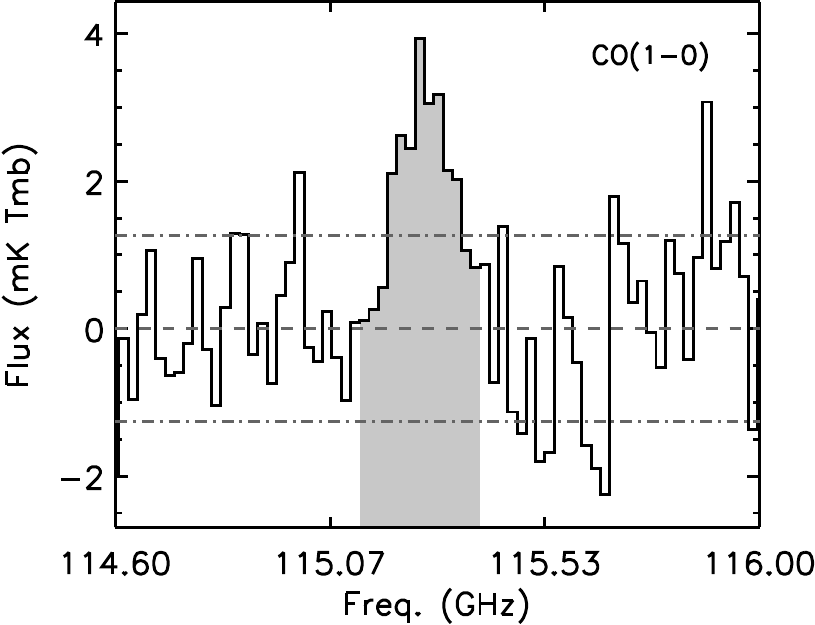} &
\includegraphics[height=4cm,angle=0,clip,trim=0.0cm 0.0cm 0cm -0.3cm]{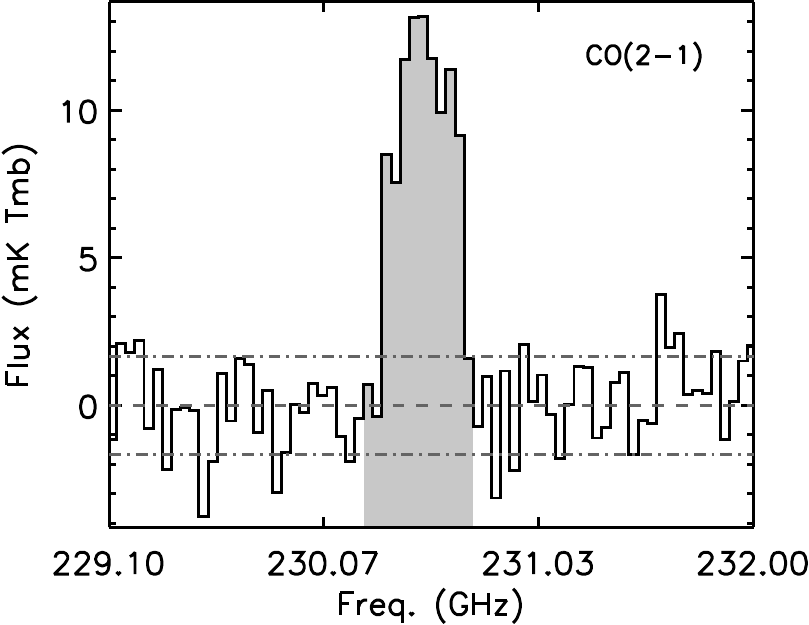}\\
\begin{turn}{90}\large \hspace{1.1cm} DETG0920 \end{turn} &
\includegraphics[height=4cm,angle=0,clip,trim=0.0cm 0.0cm 0cm 0cm]{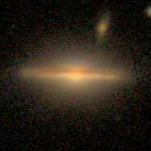} &
\includegraphics[height=4cm,angle=0,clip,trim=0.0cm 0.0cm 0cm -0.3cm]{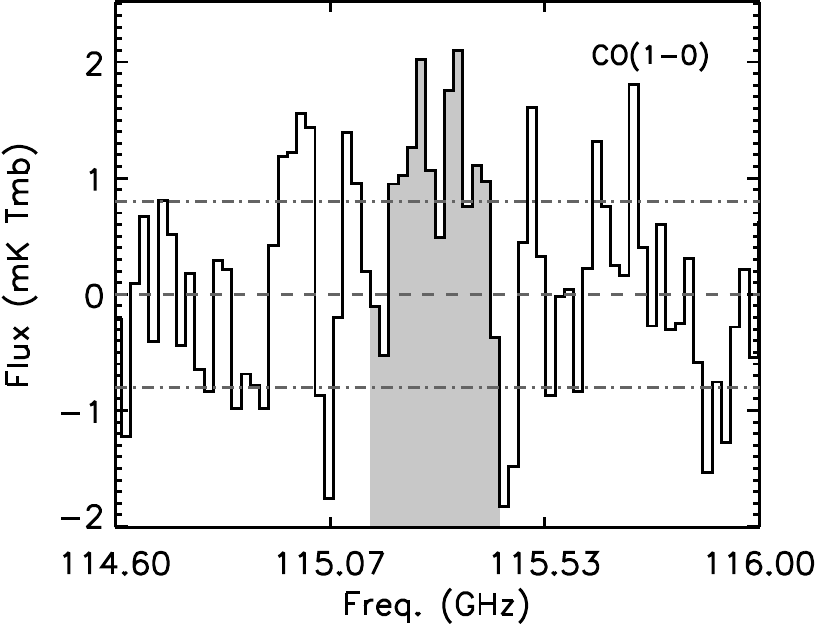} &
\includegraphics[height=4cm,angle=0,clip,trim=0.0cm 0.0cm 0cm -0.3cm]{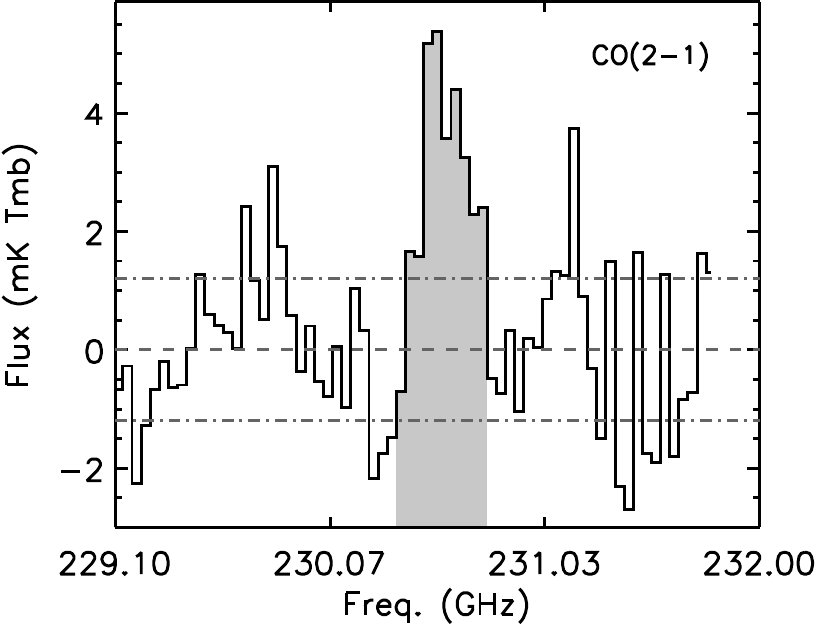}\\
\begin{turn}{90}\large \hspace{1.1cm} DETG0967 \end{turn} &
\includegraphics[height=4cm,angle=0,clip,trim=0.0cm 0.0cm 0cm 0cm]{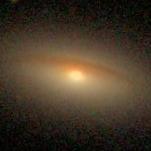} &
\includegraphics[height=4cm,angle=0,clip,trim=0.0cm 0.0cm 0cm -0.3cm]{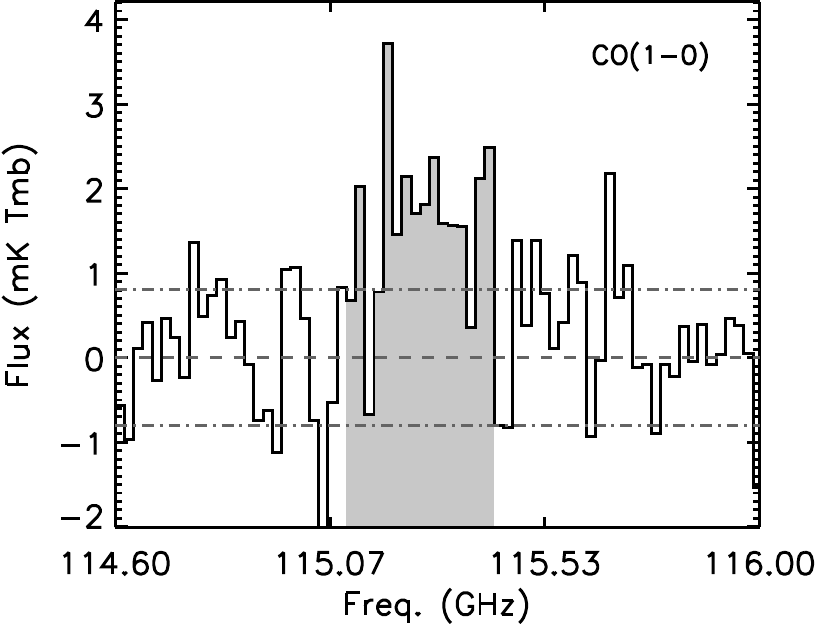} &
\includegraphics[height=4cm,angle=0,clip,trim=0.0cm 0.0cm 0cm -0.3cm]{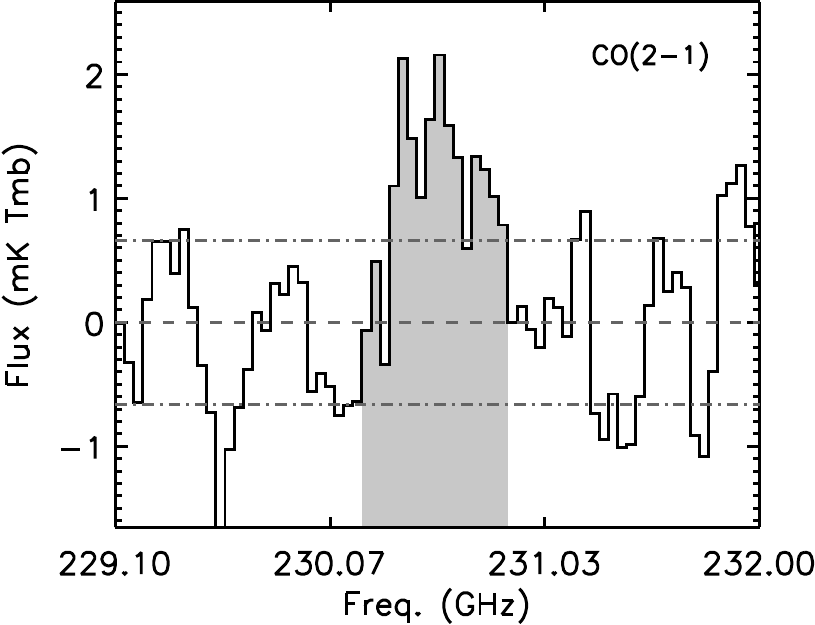} \\
\end{array}$
 \end{center}
 \caption{Three-colour (gri) optical images extracted from the SDSS (each with an identical angular size of 50$\times$50\arcsec; left column), as well as CO(1-0) and CO(2-1) IRAM-30m spectra (centre and right, respectively) for our sample DETGs (boxcar smoothed and binned to 50 \kms). The x-axis is the rest frequency for an object with expected redshift given in Table \ref{proptable}. The grey shaded region on the spectra denotes the detected line, and the dashed lines show the baseline level, and $\pm$1$\sigma$ RMS level.}
 \label{codetsfig1}
 \end{minipage}
 \end{figure*}

 \begin{figure*}
\begin{minipage}[t]{1.0\textwidth}
\begin{center}
$\begin{array}{cccc}
\begin{turn}{90}\large \hspace{1.1cm} DETG0984 \end{turn} &
\includegraphics[height=4cm,angle=0,clip,trim=0.0cm 0.0cm 0cm 0cm]{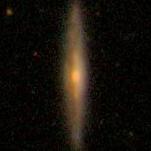} &
\includegraphics[height=4cm,angle=0,clip,trim=0.0cm 0.0cm 0cm -0.3cm]{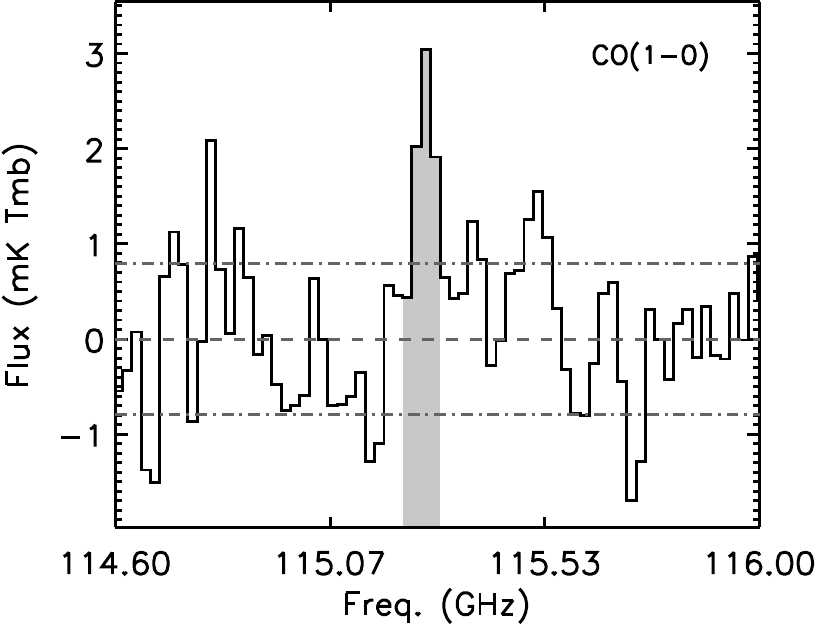} &
\includegraphics[height=4cm,angle=0,clip,trim=0.0cm 0.0cm 0cm -0.3cm]{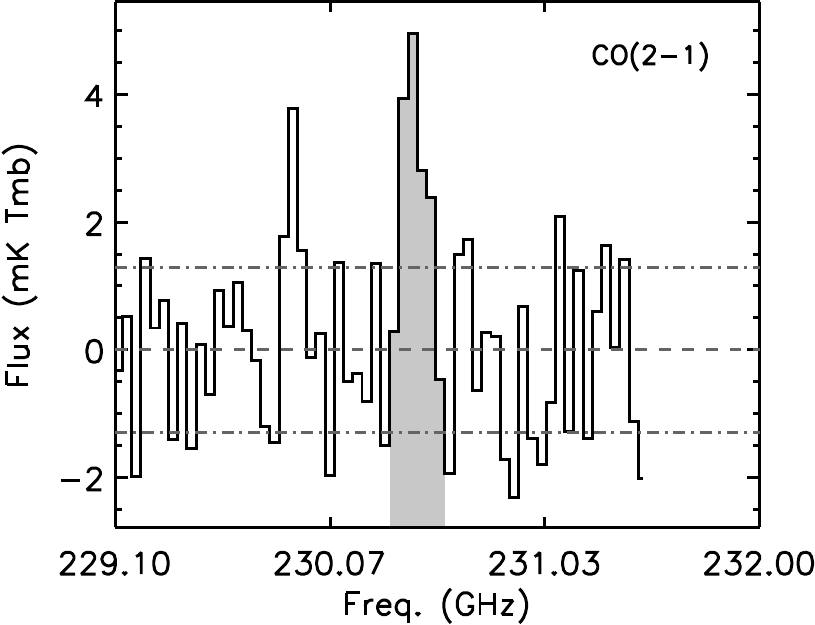}\\
\begin{turn}{90}\large \hspace{1.1cm} DETG1004 \end{turn} &
\includegraphics[height=4cm,angle=0,clip,trim=0.0cm 0.0cm 0cm 0cm]{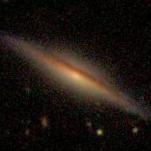} &
\includegraphics[height=4cm,angle=0,clip,trim=0.0cm 0.0cm 0cm -0.3cm]{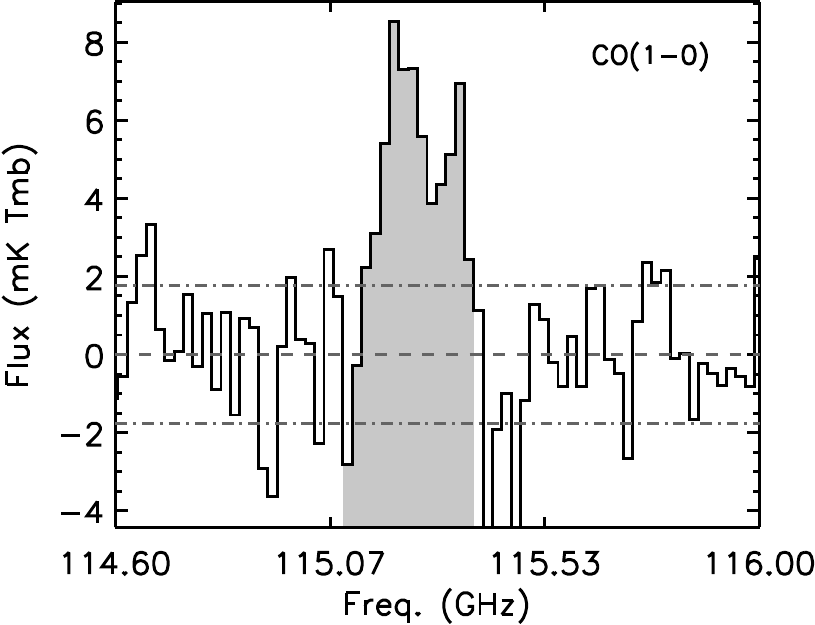} &
\includegraphics[height=4cm,angle=0,clip,trim=0.0cm 0.0cm 0cm -0.3cm]{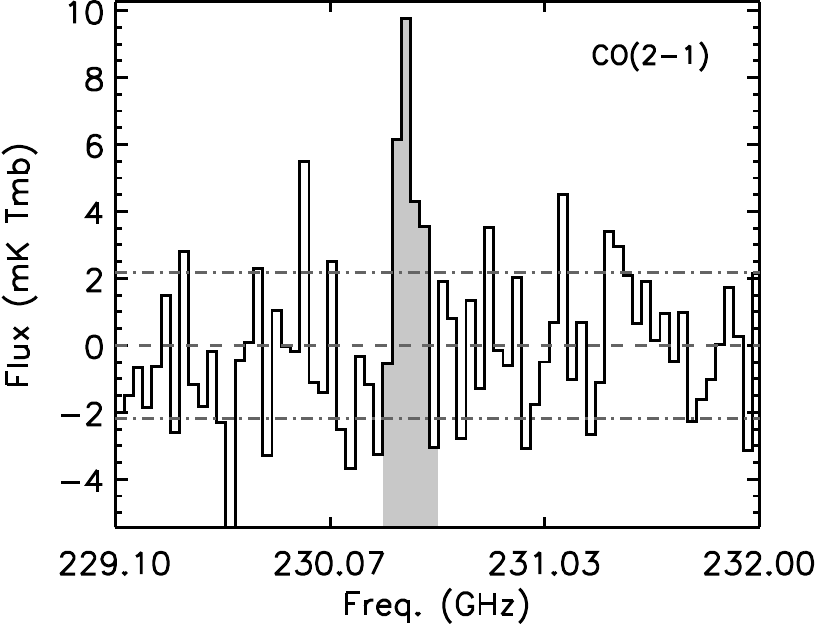}\\
\begin{turn}{90}\large \hspace{1.1cm} DETG1014 \end{turn} &
\includegraphics[height=4cm,angle=0,clip,trim=0.0cm 0.0cm 0cm 0cm]{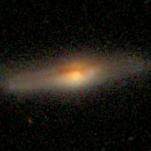} &
\includegraphics[height=4cm,angle=0,clip,trim=0.0cm 0.0cm 0cm -0.3cm]{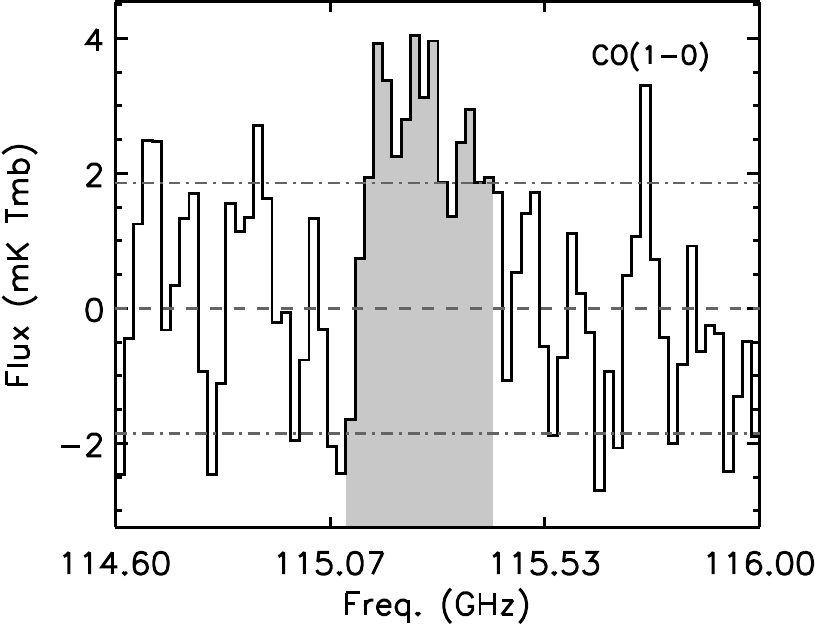} &
\includegraphics[height=4cm,angle=0,clip,trim=0.0cm 0.0cm 0cm -0.3cm]{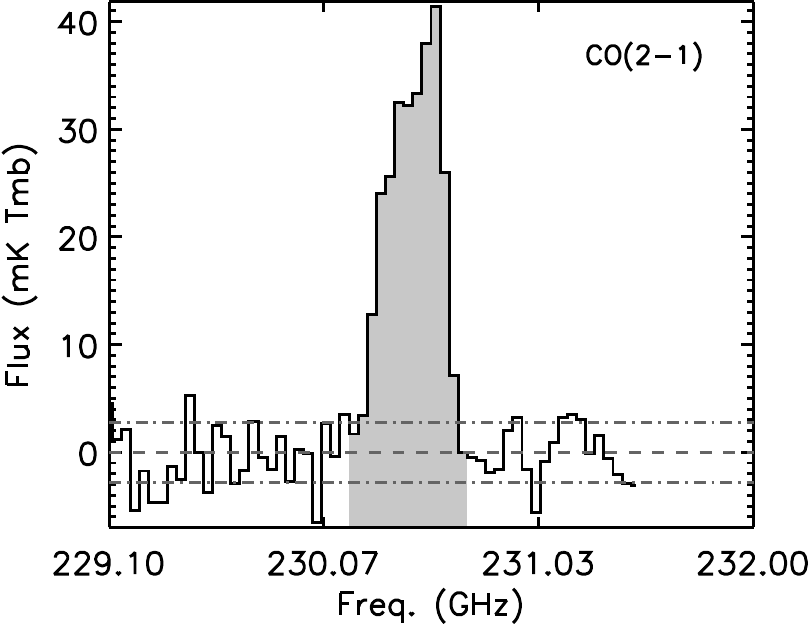}\\
\begin{turn}{90}\large \hspace{1.1cm} DETG1111 \end{turn} &
\includegraphics[height=4cm,angle=0,clip,trim=0.0cm 0.0cm 0cm 0cm]{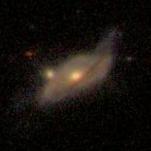} &
\includegraphics[height=4cm,angle=0,clip,trim=0.0cm 0.0cm 0cm -0.3cm]{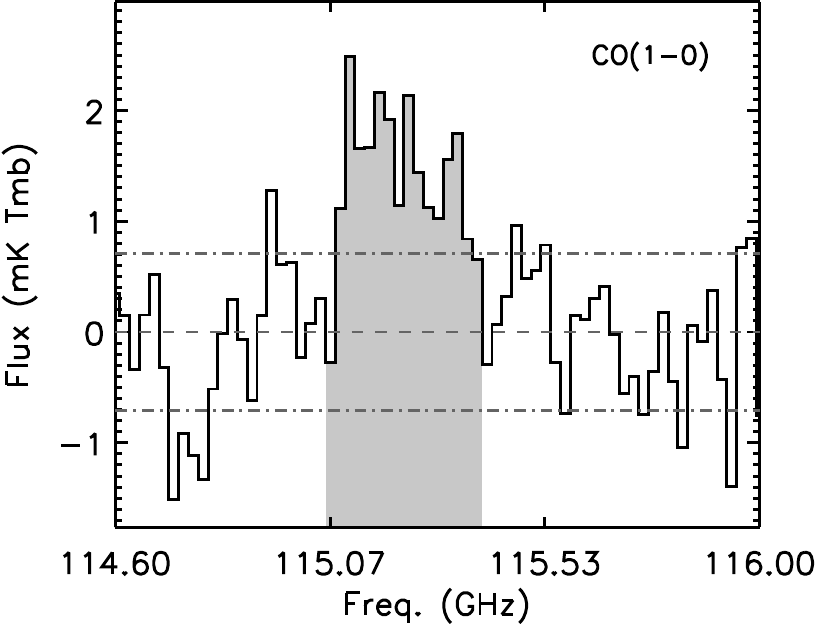} &
\includegraphics[height=4cm,angle=0,clip,trim=0.0cm 0.0cm 0cm -0.3cm]{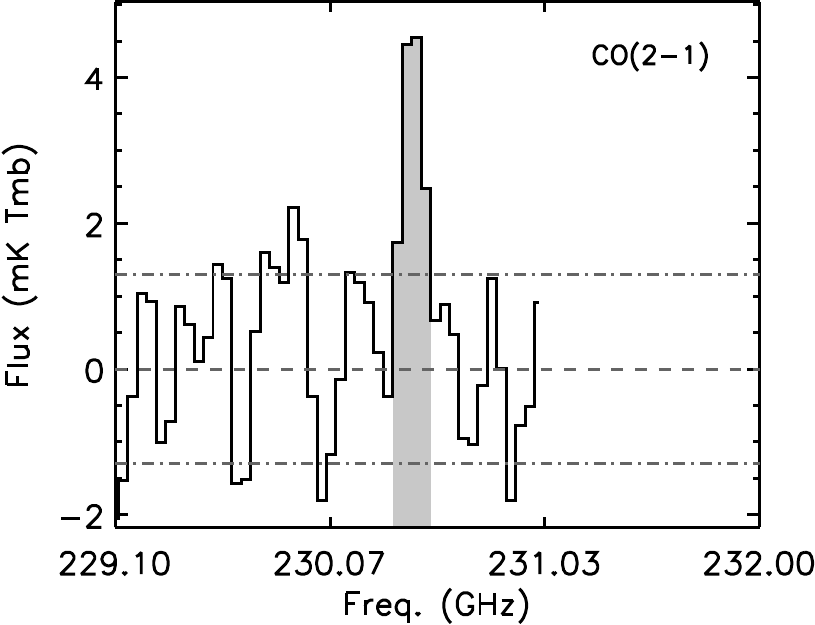}\\
\begin{turn}{90}\large \hspace{1.1cm} DETG1232 \end{turn} &
\includegraphics[height=4cm,angle=0,clip,trim=0.0cm 0.0cm 0cm 0cm]{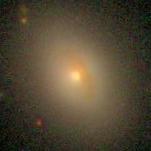} &
\includegraphics[height=4cm,angle=0,clip,trim=0.0cm 0.0cm 0cm -0.3cm]{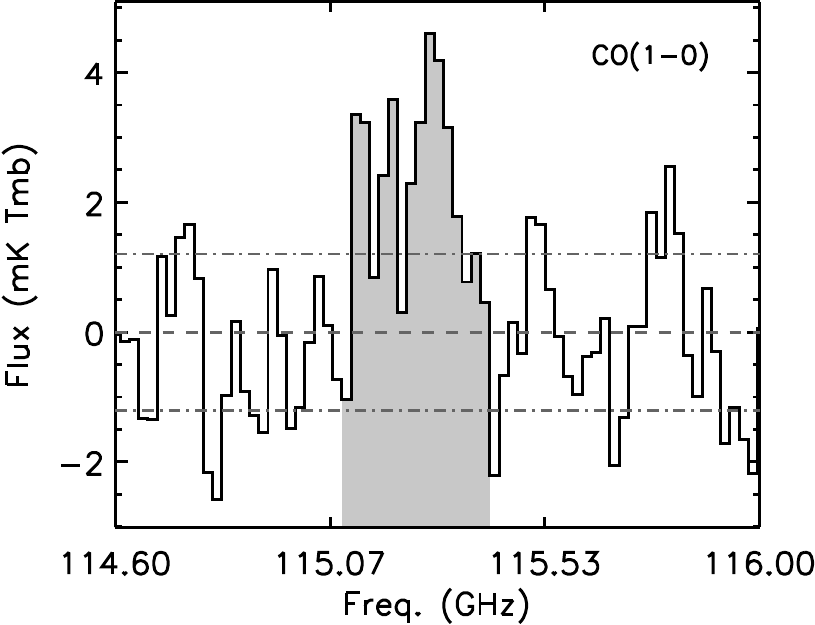} &
\includegraphics[height=4cm,angle=0,clip,trim=0.0cm 0.0cm 0cm -0.3cm]{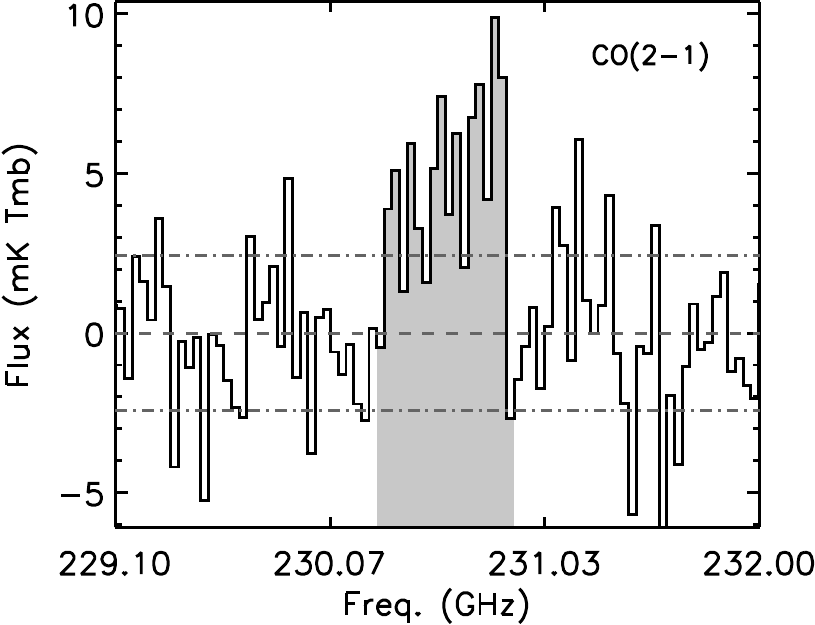}\\
\end{array}$
 \end{center}
\contcaption{}{}
 \end{minipage}
 \end{figure*}
 
\begin{figure*}
\begin{minipage}[t]{1.0\textwidth}
\begin{center}
$\begin{array}{cccc}
\begin{turn}{90}\large \hspace{1.1cm} DETG1265 \end{turn} &
\includegraphics[height=4cm,angle=0,clip,trim=0.0cm 0.0cm 0cm 0cm]{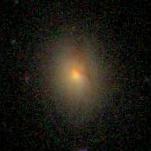} &
\includegraphics[height=4cm,angle=0,clip,trim=0.0cm 0.0cm 0cm -0.3cm]{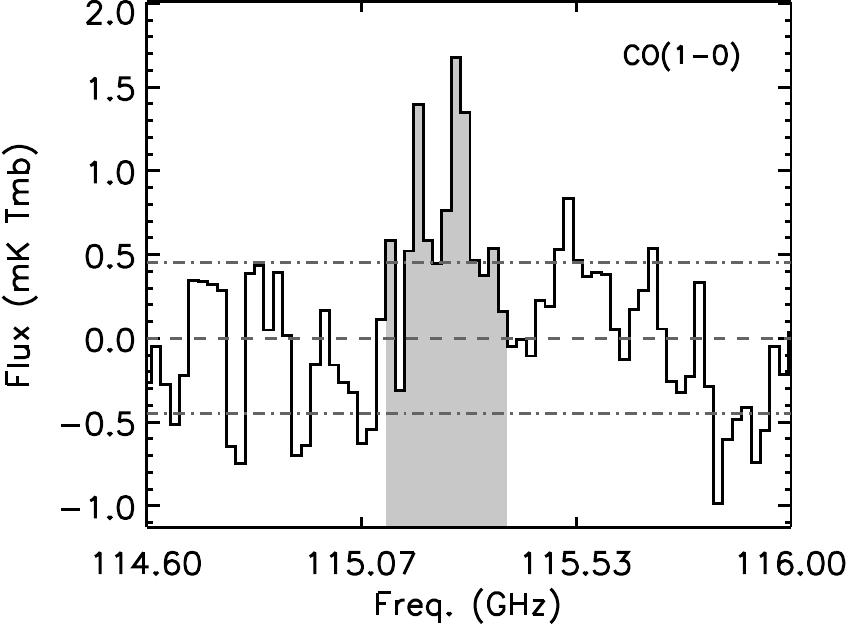} &
\includegraphics[height=4cm,angle=0,clip,trim=0.0cm 0.0cm 0cm -0.3cm]{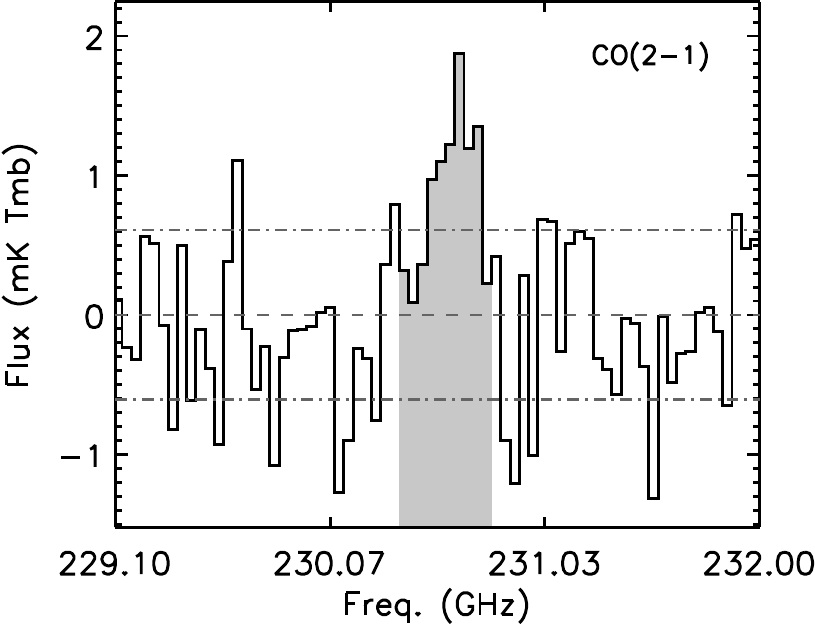}\\
\begin{turn}{90}\large \hspace{1.1cm} DETG1267 \end{turn} &
\includegraphics[height=4cm,angle=0,clip,trim=0.0cm 0.0cm 0cm 0cm]{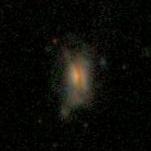} &
\includegraphics[height=4cm,angle=0,clip,trim=0.0cm 0.0cm 0cm -0.3cm]{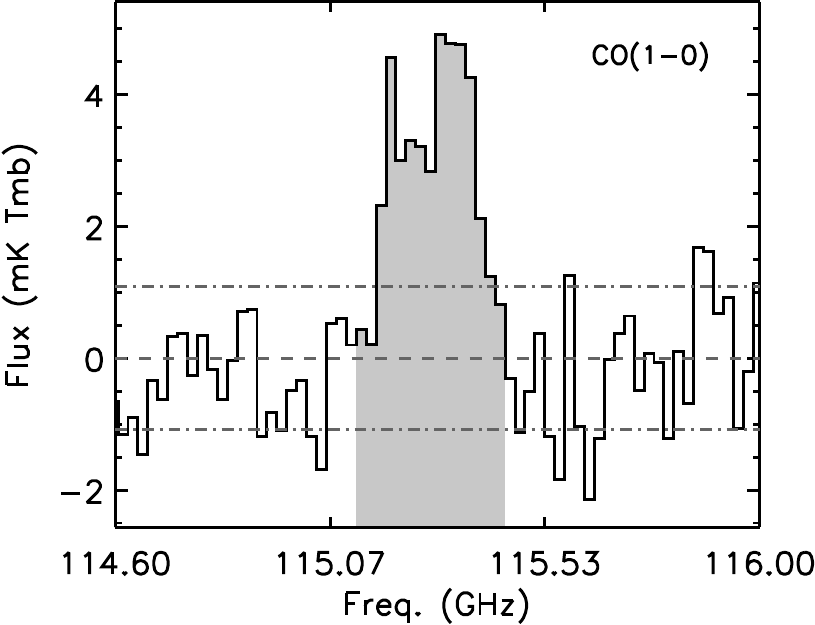} &
\includegraphics[height=4cm,angle=0,clip,trim=0.0cm 0.0cm 0cm -0.3cm]{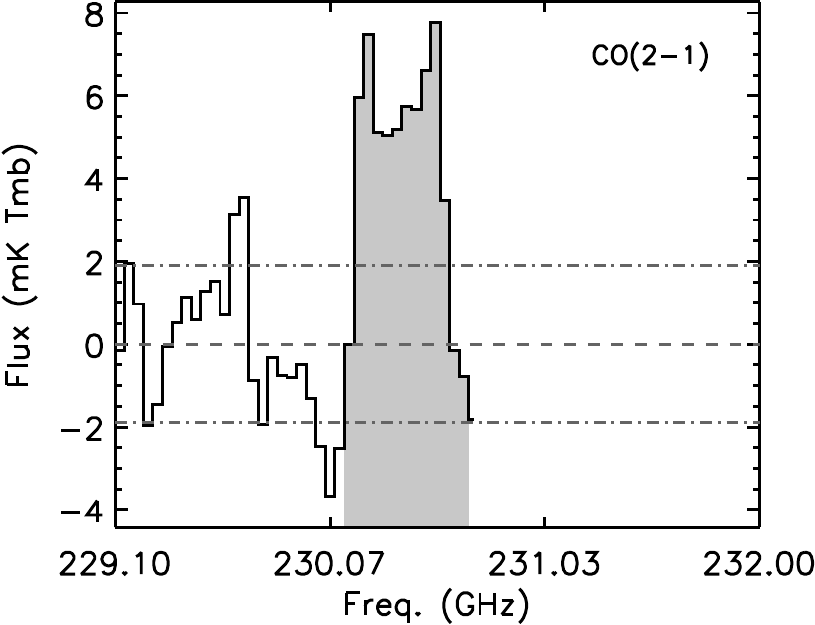}\\
\begin{turn}{90}\large \hspace{1.1cm} DETG1268 \end{turn} &
\includegraphics[height=4cm,angle=0,clip,trim=0.0cm 0.0cm 0cm 0cm]{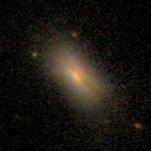} &
\includegraphics[height=4cm,angle=0,clip,trim=0.0cm 0.0cm 0cm -0.3cm]{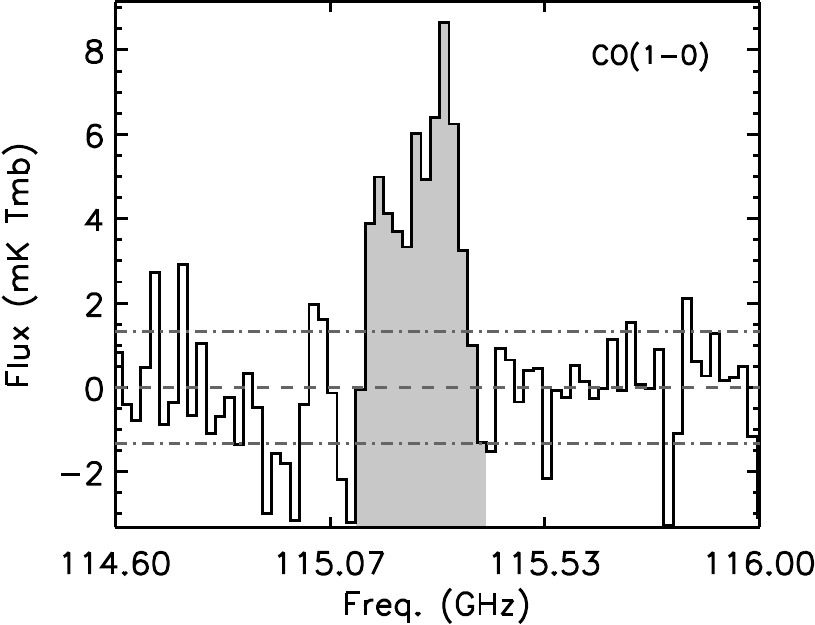} &
\includegraphics[height=4cm,angle=0,clip,trim=0.0cm 0.0cm 0cm -0.3cm]{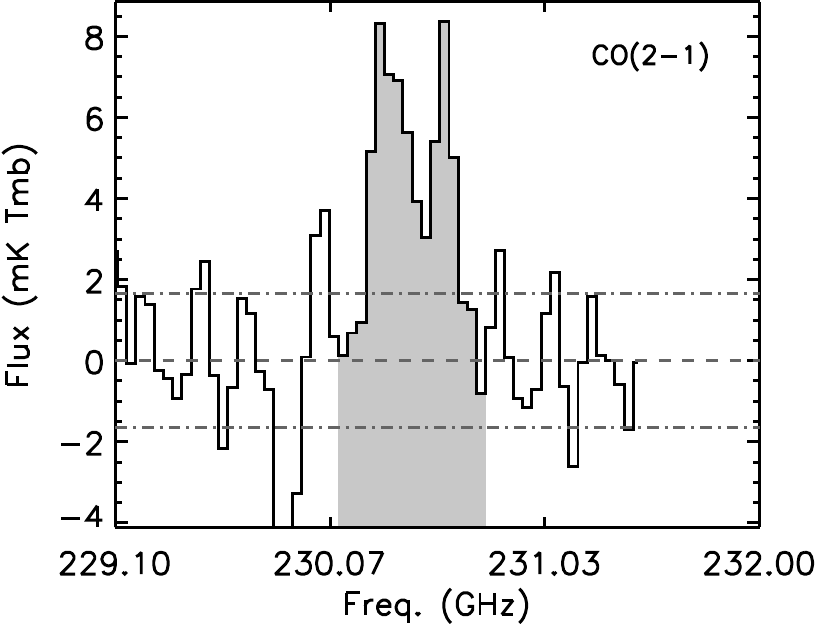}\\
\begin{turn}{90}\large \hspace{1.1cm} DETG1372 \end{turn} &
\includegraphics[height=4cm,angle=0,clip,trim=0.0cm 0.0cm 0cm 0cm]{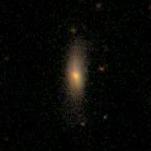} &
\includegraphics[height=4cm,angle=0,clip,trim=0.0cm 0.0cm 0cm -0.3cm]{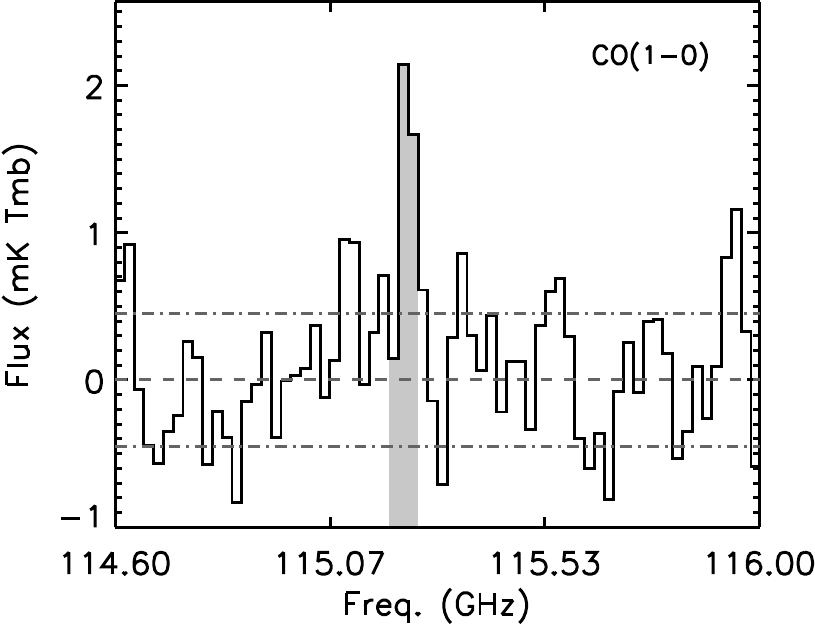} &
\includegraphics[height=4cm,angle=0,clip,trim=0.0cm 0.0cm 0cm -0.3cm]{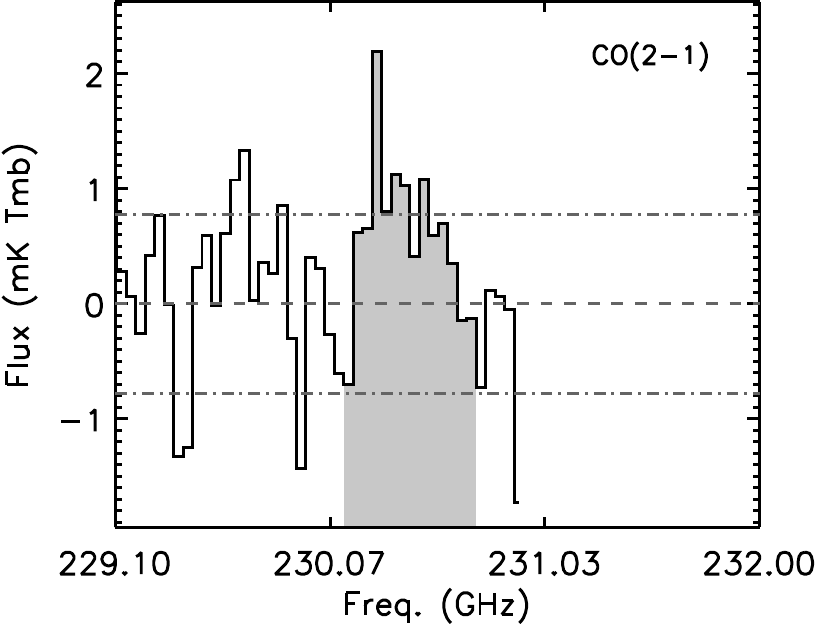}\\
\begin{turn}{90}\large \hspace{1.1cm} DETG1947 \end{turn} &
\includegraphics[height=4cm,angle=0,clip,trim=0.0cm 0.0cm 0cm 0cm]{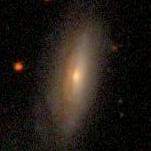} &
\includegraphics[height=4cm,angle=0,clip,trim=0.0cm 0.0cm 0cm -0.3cm]{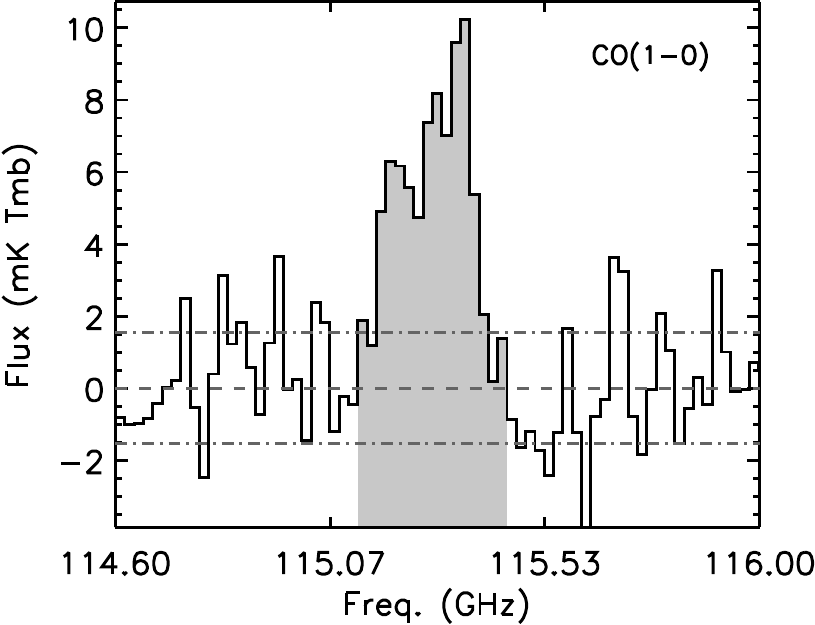} &
\includegraphics[height=4cm,angle=0,clip,trim=0.0cm 0.0cm 0cm -0.3cm]{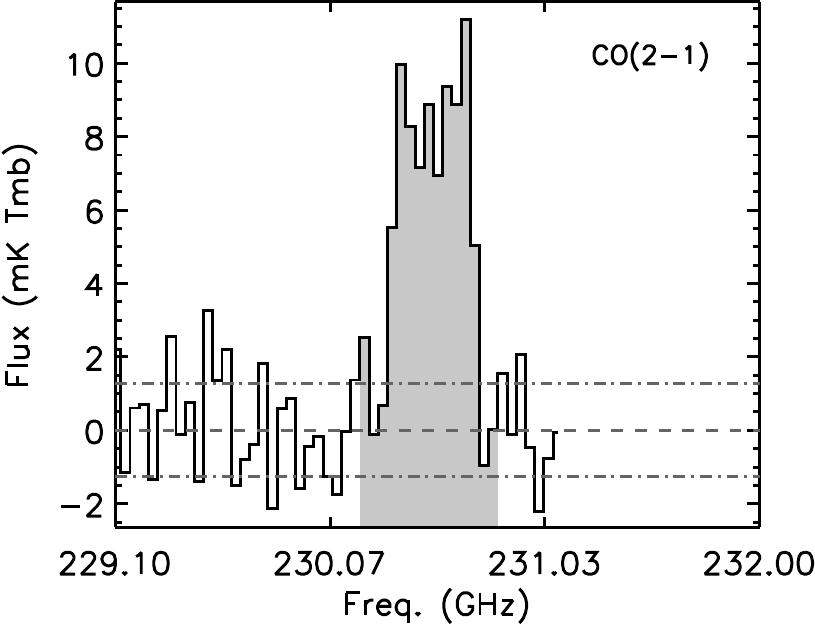}\\
\end{array}$
 \end{center}
\contcaption{}{}
 \end{minipage}
 \end{figure*}

 \begin{table*}
\caption{Observational parameters and derived molecular gas masses for the sample DETGs}
\begin{tabular*}{\textwidth}{@{\extracolsep{\fill}}l r r r r r r r r r r r r r}
\hline
ID & V$_{\rm sys}$ & Peak$_{1-0}$ & RMS$_{1-0}$ & $\Delta$V$_{1-0}$ & $\int S_{\nu}\, \delta V_{1-0}$  & Peak$_{2-1}$ &  RMS$_{2-1}$ & $\Delta$V$_{2-1}$ & $\int S_{\rm \nu}\, \delta V$$_{2-1}$ & log$_{10}$(M$_{\rm H_2}$) \\ 
  & (\kms) & (mJy) & (mJy) & (\kms) & (Jy \kms) & (mJy) & (mJy) & (\kms) & (Jy \kms) & (M$_{\odot}$) \\
 (1) & (2) & (3) & (4) & (5) & (6) & (7) & (8) & (9) & (10) & (11)\\
\hline
%Detected\\
       18 &      11700 &       32.6 &        8.3 &       410. &       12.3$\pm$       2.1 &       68.4 &       11.0 &       450. &       34.2$\pm$       3.0 &       9.64$\pm$      0.07\\
       106 &      23750 &       10.3 &        4.3 &       450. &        4.9$\pm$       1.2 &       30.5 &        5.7 &       450. &       14.8$\pm$       1.6 &       9.89$\pm$      0.11\\
       231 &       8378 &       20.4 &        7.4 &       358. &        7.8$\pm$       1.7 &      110.0 &       13.0 &       364. &       41.0$\pm$       3.2 &       9.19$\pm$      0.10\\
       920 &       7550 &        9.5 &        4.7 &       391. &        3.3$\pm$       1.2 &       40.0 &        9.4 &       286. &       12.0$\pm$       2.0 &       8.65$\pm$      0.15\\
       967 &       7936 &       14.2 &        4.7 &       450. &        7.7$\pm$       1.3 &       14.6 &        5.2 &       450. &        7.3$\pm$       1.4 &       9.05$\pm$      0.07\\
       984 &      10957 &       17.5 &        4.7 &       143. &        2.5$\pm$       0.7 &       31.6 &       10.2 &       173. &        5.1$\pm$       1.7 &       8.88$\pm$      0.13\\
      1004 &      11675 &       43.4 &       10.4 &       416. &       18.8$\pm$       2.6 &       53.1 &       17.1 &       168. &        9.4$\pm$       2.8 &       9.83$\pm$      0.06\\
      1014 &       7150 &       22.4 &       11.0 &       450. &       11.9$\pm$       2.9 &      310.0 &       22.0 &       369. &      116.7$\pm$       5.4 &       9.21$\pm$      0.11\\
      1111 &      18450 &       12.7 &        4.2 &       450. &        7.3$\pm$       1.2 &       36.7 &       10.2 &       163. &        6.1$\pm$       1.7 &       9.84$\pm$      0.07\\
      1232 &       7650 &       20.9 &        7.1 &       450. &        9.9$\pm$       1.9 &       53.1 &       19.2 &       425. &       27.1$\pm$       4.5 &       9.17$\pm$      0.08\\
      1265 &      11903 &        6.6 &        2.7 &       378. &        2.6$\pm$       0.7 &       12.0 &        4.8 &       299. &        3.9$\pm$       1.1 &       8.98$\pm$      0.11\\
      1267 &      23419 &       28.7 &        6.4 &       450. &       14.2$\pm$       1.8 &       52.5 &       14.9 &       438. &       24.6$\pm$       3.6 &      10.31$\pm$      0.05\\
      1268 &      10850 &       39.2 &        7.8 &       412. &       16.7$\pm$       2.0 &       55.1 &       12.9 &       450. &       26.9$\pm$       3.5 &       9.71$\pm$      0.05\\
      1372 &      18530 &       13.8 &        2.7 &       104. &        1.5$\pm$       0.4 &        8.2 &        6.1 &       433. &        3.4$\pm$       1.6 &       9.17$\pm$      0.11\\
      1947 &      10283 &       52.1 &        9.1 &       450. &       25.3$\pm$       2.4 &       78.1 &       10.0 &       450. &       36.1$\pm$       2.7 &       9.84$\pm$      0.04\\
      \hline
   %   Non det.\\
      1030 & -- &  -- & 3.2 &  -- & -- & -- & 4.2 &  -- &  -- & $<$8.94\\ 
      1362 & -- &  -- & 5.7 &  -- & -- & -- & -- &  -- &  -- & $<$10.19\\ 
         \hline
\end{tabular*}
\parbox[t]{1\textwidth}{ \textit{Notes:}  Column 1 lists the DETG ID number of each source. Column 2 shows the systemic velocity (in the local standard-of-rest frame) derived from fitting the CO(1-0) and CO(2-1) spectra. Column 3 lists the peak CO(1-0) line flux for each object, and column 4 the RMS noise reached in the observation (in $\approx$50\kms channels). Column 5 shows the CO(1-0) linewidth, and column 6 the integrated intensity of the CO(1-0) line. Columns 7-10 show the same properties as columns 3-6, but derived from the CO(2-1) line. Dashes indicate sources where 1mm observations were not performed. Column 11 shows the logarithm of the H$_2$ mass for each source, derived using Equation \protect \ref{co2h2}. For the two non detected sources Column 11 lists the 3$\sigma$ upper limit to the H$_2$ mass derived assuming a 200\kms\ velocity width for the line. }
\label{obstable}
\end{table*}

 \begin{table*}
\caption{Derived parameters for the sample DETGs}
\begin{tabular*}{0.8\textwidth}{@{\extracolsep{\fill}}l r r r r r}
\hline
ID &  M$_{\rm H_2}$/M$_{\rm d}$ & M$_{\rm H_2}$/M$_{\rm *}$ & M$_{\rm H_2}$/M$_{\rm \hi}$ & M$_{\rm gas}$/M$_{\rm d}$ & M$_{\rm gas}$/M$_{\rm *}$ \\
  & & (\%) & & &(\%) \\
 (1) & (2) & (3) & (4) & (5) & (6) \\
\hline
        18 &        172$\pm$        29 &        6.2$\pm$       1.1 & -- & $>$ 172 & $>$ 6.2\\
       106 &         61$\pm$        15 &        5.4$\pm$       1.3 &        1.3$\pm$      0.35 &        107$\pm$       15. &        10.$\pm$       1.3\\
       231 &         34$\pm$         7 &        4.6$\pm$       1.0 & -- & $>$  34 & $>$ 4.6\\
       920 &         23$\pm$         8 &        0.7$\pm$       0.2 &        0.9$\pm$      0.33 &         50$\pm$        8. &         1.$\pm$       0.2\\
       967 &         28$\pm$         4 &        0.8$\pm$       0.1 &        1.9$\pm$      0.39 &         44$\pm$        5. &         1.$\pm$       0.1\\
       984 &         28$\pm$         8 &        1.2$\pm$       0.4 &        0.2$\pm$      0.05 &        200$\pm$        8. &         9.$\pm$       0.4\\
      1004 &         51$\pm$         7 &       10.2$\pm$       1.4 &        0.5$\pm$      0.06 &        164$\pm$        7. &        33.$\pm$       1.4\\
      1014 &         60$\pm$        14 &        4.0$\pm$       1.0 &        0.2$\pm$      0.06 &        315$\pm$       15. &        21.$\pm$       1.0\\
1030  &  $<$34  & $<$0.02 & -- & -- & -- \\
      1111 &         45$\pm$         7 &        6.0$\pm$       1.0 &        0.3$\pm$      0.04 &        209$\pm$        7. &        28.$\pm$       1.0\\
      1232 &        145$\pm$        27 &        1.0$\pm$       0.2 &        0.4$\pm$      0.09 &        518$\pm$       27. &         3.$\pm$       0.2\\
      1265 &         33$\pm$         8 &        1.6$\pm$       0.4 &        0.1$\pm$      0.02 &        402$\pm$        8. &        19.$\pm$       0.4\\
      1267 &        227$\pm$        28 &       23.4$\pm$       2.9 &        2.0$\pm$      0.25 &        344$\pm$       29. &        35.$\pm$       2.9\\
      1268 &        467$\pm$        55 &        7.5$\pm$       0.9 &        1.8$\pm$      0.23 &        730$\pm$       56. &        12.$\pm$       0.9\\
1362  &  $<$90  & $<$0.15 & -- & -- & -- \\
      1372 &         23$\pm$         5 &        3.2$\pm$       0.8 &        0.1$\pm$      0.02 &        376$\pm$        6. &        52.$\pm$       0.8\\
      1947 &        112$\pm$        10 &        8.7$\pm$       0.8 &        0.4$\pm$      0.04 &        373$\pm$       11. &        29.$\pm$       0.8\\              \hline
\end{tabular*}
\parbox[t]{0.8\textwidth}{ \textit{Notes:} Column 1 lists the DETG ID number of each source. Columns 2, 3 and 4 contain the ratio of the molecular hydrogen mass to the dust mass, stellar mass and \hi\ mass, respectively, while Columns 5 and 6 contain the ratio of the total gas mass (\hi + H$_2$) to the dust mass, and the stellar mass respectively. Dashes indicate objects where observations were not available, where the lower limit to the gas-to-dust and gas-to-stellar mass ratios are given. }
\label{ratiotable}
\end{table*}

\section{Results}
\label{results}
\subsection{Line detections}

We formally detect the CO($J$=1-0) and CO($J$=2-1) lines in 15 of our 17 DETGs. The CO spectra for the detections are presented along with SDSS 3-colour images of the sources in Figure \ref{codetsfig1}. Table \ref{obstable} lists the properties of the detected lines, and upper limits for the non-detections. We do not discuss the non-detections further, as the observations we have are not deep enough to be constraining (but we do include these objects in future tables for completeness). In all cases with detected lines the redshifts of the two CO transitions are consistent with one another, and with the optical spectroscopic redshift of the system (Table \ref{proptable}).

The velocity width of a CO line in a galaxy depends both on the rotation curve of the galaxy, and the relative extent of the emitting components. Thus real velocity width differences between different CO transitions may indicate that the emitting gas is concentrated differently or that the smaller beam at higher frequencies is missing some gas.  The velocity widths of the detected lines in our target DETGs are consistent in most cases, but we discuss individual cases where they differ below. The velocity linewidths found for our sources (upto 450 \kms) are consistent with the expectation for ETGs of this mass from the CO Tully-Fisher relation; e.g. \cite{Davis:2011bg}.

In DETG1372 the CO(1-0) line is very much narrower than the CO(2-1) line. This is hard to explain physically, and noting the low significance of both line detections ($<$4$\sigma$) we suspect that one or both may be spurious detections. Treating the H$_2$ mass derived for this source as an upper limit in our analysis does not affect our conclusions, as this source is extremely \hi\ dominated.

DETG0920, DETG1004, DETG1014 and DETG1111 have a significantly smaller CO(2-1) velocity width than the CO(1-0) velocity width, and thus either the CO(2-1) emitting gas is more centrally concentrated, or the smaller beam of the telescope at 230 GHz causes us to miss CO(2-1) emission at higher velocities. These objects do not seem to share any distinctive physical properties (e.g. their morphologies, starburst ages etc. seem to span the whole range found in the underlying population; \citealt{2012MNRAS.423...59S}). {  \cite{2013MNRAS.429..534D} studied the extent of the gas in \atlas\ ETGs, and found that the gas typically extends to half the effective radius. The objects with discrepant CO(2-1) velocity widths have optical effective radii of between 15-25\arcsec, suggesting gas extents of 7.5-12.5\arcsec. As these objects are substantially more disturbed than a typical \atlas\ object, their gas may be even more extended, making it feasible that an 11\arcsec\ beam could miss some flux (see the left column of Figure \ref{codetsfig1}). Beam effects are thus the more likely cause of this discrepancy.}

In principle, the ratio of the CO(1--0) and CO(2--1) lines can be used to estimate the excitation temperature of the molecular gas, but the significantly smaller beam size of the IRAM-30m at the higher frequency of CO(2--1) acts as a systematic error in such a measurement.
Thus the line ratio we observe is affected by both the excitation temperature and the spatial distribution of the gas.
 If the observed CO emission was to fill the telescope beam at both frequencies (and the CO is not sub-thermally excited) we would expect a line ratio of one (measured in antenna temperature units). However, if the CO emission is compact compared to the beam then the measured intensity in the CO(2--1) line should be larger by up to a factor of 4 (as the beam at such frequencies covers a 4 times smaller area). 
  In all but one of our objects the CO(2--1)/CO(1--0) antenna temperature ratio is less than 4, suggesting the total CO extent is small compared to our beam.  It should be noted however that sub-thermal excitation of CO lines would decrease the intensity of the higher frequency transition \citep[e.g.][]{1992A&A...264..433B}, so a firm statement about the extent of the gas is not possible without resolved data.  DETG1014 is the exception, having a CO(2--1)/CO(1--0) ratio of $\approx$7.3, suggesting a high excitation temperature may be required in this object.

\subsection{Gas masses}
\label{gasmasses}

\subsubsection{\hi}
The galaxies in our sample were searched for 21cm \hi\ emission at the Arecibo telescope, as detailed in {Allison et al., in prep.} (paper two of this series). All but two of our objects were detected, and in this paper we use the \hi\ masses (M$_{\rm \hi}$) as described in {Allison et al.} (see that work for full details). NGC 4797 (DETG 1232) already has an \hi\ mass measurement in the ALFALFA 40\% catalogue \citep{2011AJ....142..170H}, which reports that M$_{\rm \hi}$ = 3.74 $\pm$ 0.54 $\times$10$^{9}$ M$_{\odot}$, and we use this value here.

\subsubsection{H$_2$}
We estimate H$_2$ gas masses for our CO detections in the standard manner, using the following equation

\begin{equation}
M_{H_2} = 2m_H \frac{\lambda^2}{2 k_b} X_{\rm CO} D_L^2 \int{S_v \delta V},
\end{equation}
\noindent where $m_H$ is the mass of a hydrogen atom, $\lambda$ is the wavelength, $k_b$ is Boltzmann's constant, $D_L$ is the luminosity distance, $\int{S_v \delta V}$ is the integrated CO flux density and  X$_{\rm CO}$ is your CO-to-H$_2$ conversion factor of choice. 
For CO(1-0) this formula reduces to

\begin{equation}
M_{H_2} = 3.93\times10^{-17} X_{\rm CO} \left(\frac{D_L}{\mathrm{Mpc}}\right)^2 \left(\frac{\int{S_v \delta V}}{\mathrm{Jy\,km\,s^{-1}}}\right).
\label{co2h2}
\end{equation}

Here we use a galactic X$_{\rm CO}$ factor of 3$\times$10$^{20}$ cm$^{-2}$ (K \kms)$^{-1}$, equivalent to $\alpha_{\rm CO}$= 4.4 \msun (K \kms pc$^2$)$^{-1}$ \citep{Dickman:1986jz}. An X$_{\rm CO}$ more appropriate for low-metallicity gas would produce even higher H$_2$ masses. We return to discuss this assumption later, in the light of our derived gas-to-dust ratios (see Section \ref{discuss_xco}).

We find that the detected galaxies are very gas rich, having H$_2$ masses between 4$\times$10$^{8}$ and 2$\times$10$^{10}$ \msun. The value derived for each object is listed in Table \ref{obstable}. We discuss the gas rich nature of these objects further in Section \ref{h2rich}.

\subsection{Gas-to-dust and gas-to-stellar mass ratios}
\label{g2d_g2s}

 \begin{figure}
\begin{center}
\includegraphics[width=0.45\textwidth,angle=0,clip,trim=0.0cm 0.0cm 0cm 0.0cm]{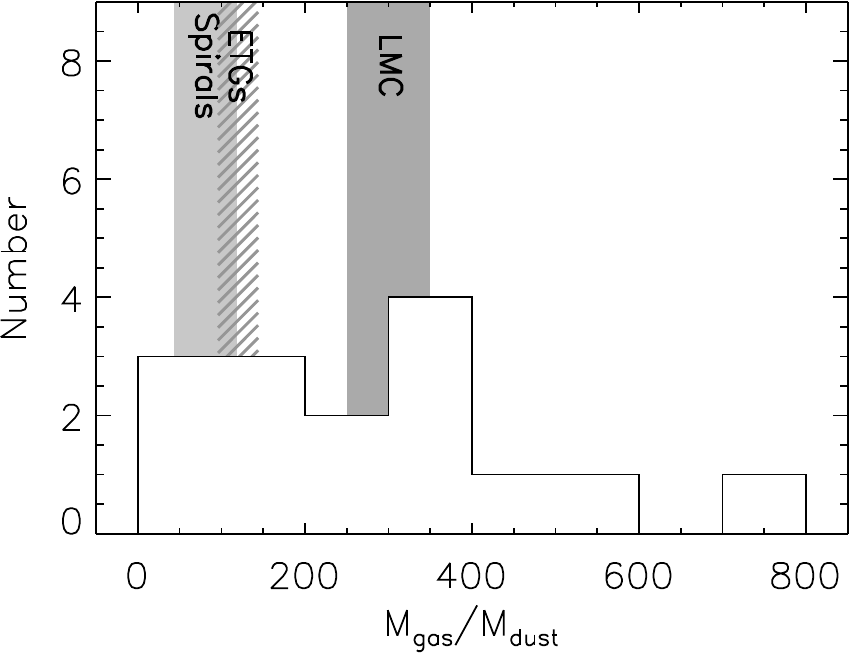}
 \end{center}
\caption{A histogram of the measured gas-to-dust ratios in our sample DETGs, taken from Table \ref{ratiotable}. Also shown as shaded areas are the typical gas-to-dust ratio found in massive spiral galaxies, and early type galaxies (from the \emph{Herschel} Reference Survey; \citealt{2012ApJ...748..123S}), and the measured gas-to-dust ratio in the Large Magellanic Cloud (LMC; \citealt{2010A&A...518L..74R}).}
\label{g2dfig}
 \end{figure}
 
In Table \ref{proptable} we presented dust and stellar mass estimates for these objects, derived from multi-wavelength SED fitting, following K13. In Table \ref{ratiotable} we present the molecular gas-to-dust and stellar mass ratios. We also combine the molecular gas masses from this work with atomic gas masses from {Allison et al.} to tabulate molecular-to-atomic gas ratios and total gas-to-dust (and gas-to-stellar mass) ratios for our sample. 
Our galaxies break into two rough groups, with $\approx$1/3 of the sample being very molecular gas rich (i.e. M$_{\rm H_2}$/M$_{\rm \hi}$ $>$ 1; DETG106, 967, 1267 and 1268) while the rest are atomic gas dominated. 

We find molecular-to-stellar mass ratios (M$_{\rm H2}$/M$_{*}$) of between 0.7 and 24\%, and total gas-to-stellar mass ratios (M$_{\rm gas}$/M$_{*}$) between 1 and 52\%.
The gas-to-dust ratios (M$_{\rm gas}$/M$_{\rm d}$) of our sample vary widely, between values of $\approx$50 and 750 (see Figure \ref{g2dfig}).

\subsection{Gas phase metallicity estimates}
\label{gasphasez}
{ The wide range of gas-to-dust ratios found in our objects imply a large range of gas-phase metallicities. If we assume the same fraction of the major condensable elements are in solid form as in the Milky Way (MW), we can use the results of \cite{2007ApJ...663..866D} to estimate the gas phase metallicity:}
\begin{equation}
{\rm 12+\log_{10}(O/H)} = 12 + \log_{10}\left(\frac{4.57088\times10^{-2}}{\rm M_{\rm gas}/{M_{\rm dust}}}\right).
\end{equation}

{ The numerical factor in the above equation is set so that at a gas-to-dust ratio of 100, one would predict solar metallicity gas ${\rm 12+\log_{10}(O/H)=8.66}$ \citep{2004A&A...417..751A}. 
Using the above equation we find that our objects are predicted to have gas phase metallicities between a tenth and twice solar metallicity (${\rm 12+\log_{10}(O/H)}$ = 7.8 to 9.0).
We discuss this result further in Section \ref{prog_estimate}.}

\section{Discussion}
\label{discuss}

\subsection{Gas masses in DETGs}
\label{h2rich}

The bulge-dominated DETGs studied in this work are very gas rich. Each detected object has a molecular gas mass $\gtsimeq4\times10^8$ \msun (see Section \ref{gasmasses}), and the most gas rich galaxy has $\approx$2$\times$10$^{10}$ \msun of molecular gas. In comparison the \atlas\ survey \citep{2011MNRAS.413..813C}, which included a complete and volume limited exploration of the gas content of 260 local ETGs \citep{2011MNRAS.414..940Y,2012MNRAS.422.1835S}, only contained three galaxies with $>$10$^9$ \msun\ of molecular gas. 

The selection criteria we used to define our sample likely explain the gas-rich nature of these objects. The galaxies show large dust lanes in shallow SDSS imaging and are the brightest DETGs in the sky in the far infrared, already suggesting they must contain reasonable amounts of ISM material. The selected DETGs are a reasonably rare population of very gas rich bulge-dominated galaxies, which are the `tip of the iceberg' for a larger population of gas-hosting ETGs, such as that explored by \atlas. They are a good sample for tracing the minor merger process however, as the optical disturbances (see \citealt{2012MNRAS.423...49K,2012MNRAS.423...59S}) and large remaining gas masses both suggest the merger happened recently. 

\subsubsection{X$_{\rm co}$}
\label{discuss_xco}
The molecular masses used here assume a Galactic $X_{\rm CO}$ factor (the conversion between CO flux and H$_2$). $X_{\rm CO}$ has been shown to vary as a function of metallicity \citep[e.g. ][]{1995ApJ...448L..97W,2008ApJ...686..948B,2011ApJ...737...12L,2013ApJ...777....5S} and in strong starbursts \citep[e.g.][]{1997ApJ...478..144S,1998ApJ...507..615D}. See \cite{2013ARA&A..51..207B} for a review of this issue. 

Our objects are not starbursts, and do not have suitable strong emission lines that allow us to directly estimate the ISM metallicity. However, as shown in Section \ref{gasphasez} the gas-phase metallicity may be sub-solar in a large number of these objects (see Section \ref{prog_estimate}). 
Calibrations between gas-phase metallicity and $X_{\rm CO}$ do exist, but we are unable to apply these here as our proxy for metallicity (the gas-to-dust ratio) itself depends on $X_{\rm CO}$. We proceed from this point using the molecular gas masses estimated from a fixed, solar metallicity $X_{\rm CO}$, but remain aware that we may be \textit{underestimating} the amount of molecular gas present in these systems (by a factor of up to 20 in some cases; \citealt{2007ApJ...658.1027L}). Total gas masses are better constrained, especially in the $\approx$2/3 of our sample galaxies that are atomic gas dominated (see below). If substantial amounts of gas exists outside our telescope beam (which covers the inner $\approx$22\arcsec\ of these galaxies) then this would also lead to underestimating the molecular gas mass present.

\subsubsection{Molecular vs atomic gas}

In Section \ref{g2d_g2s} we presented the molecular-to-atomic gas ratios in our objects, finding that $\approx$1/3 of the sample are very molecular gas rich (i.e. M$_{\rm H_2}$/M$_{\rm \hi}$ $>$ 1; DETG106, 967, 1267 and 1268) while the rest are atomic gas dominated. The molecular-to-atomic gas ratio does not correlate with the gas-to-dust ratio in this sample of objects, or with their redshift (as would be expected if beam effects cause us to miss CO emitting gas in the nearest objects).  
In the \atlas\ survey, molecule dominated galaxies also made up $\approx$1/3 of the population, { with a similar median molecular-to-atomic gas ratio ($\approx$0.3-0.5). A Mann Whittingly U test is unable find difference in the distributions of molecular-to-atomic gas ratios}, suggesting that in our DETGs the gas is in a reasonably similar physical state to that in the \atlas\ objects (despite the large total gas mass present). 

\subsubsection{Gas to stellar mass fractions}
As discussed previously our DETGs contain large amounts of gas and thus, despite being very massive objects, their gas masses are reasonable proportions of the total baryonic mass of the system. We find (see Table \ref{ratiotable}) molecular-to-stellar mass ratios (M$_{\rm H2}$/M$_{*}$) of between 0.7 and 24\%, and total gas-to-stellar mass ratios (M$_{\rm gas}$/M$_{*}$) as high as 52\% in our most extreme object (DETG1372). 
The recent merger these objects have experienced causes these objects to be as gas rich as a typical Sb or Sc type spiral galaxy in some cases \citep[e.g.][]{1991ARA&A..29..581Y,2014A&A...564A..66B}.

\subsection{Dust \& gas accretion in minor mergers}
\label{prog_estimate}

As discussed in detail above, the galaxies in our sample are expected to have undergone a recent minor merger. 
For optically red, massive bulge-dominated galaxies like these it is likely that the massive progenitor was gas poor, and the minor companion brought in the vast majority of the gas and dust into the system (see e.g. \citealt{2011MNRAS.417..882D} for a dynamical study showing an external origin { for the gas in $>$50\% of such galaxies}, and \citealt{2014MNRAS.444.3408Y} for an in-depth discussion of this hypothesis in the context of the colours of local ETGs). In addition, if a substantial gas mass were present in these objects before the merger, then the acquisition of more gas would likely cause a build-up of gas in the galaxy's centre, triggering a nuclear starburst. We do not see any evidence for such nuclear starbursts in our objects (which would be obvious in e.g. the SDSS spectra), suggesting once more that these objects were gas poor before their recent merger. 

 \begin{figure}
\begin{center}
\includegraphics[width=0.48\textwidth,angle=0,clip,trim=0.0cm 0.0cm 0cm 0.0cm]{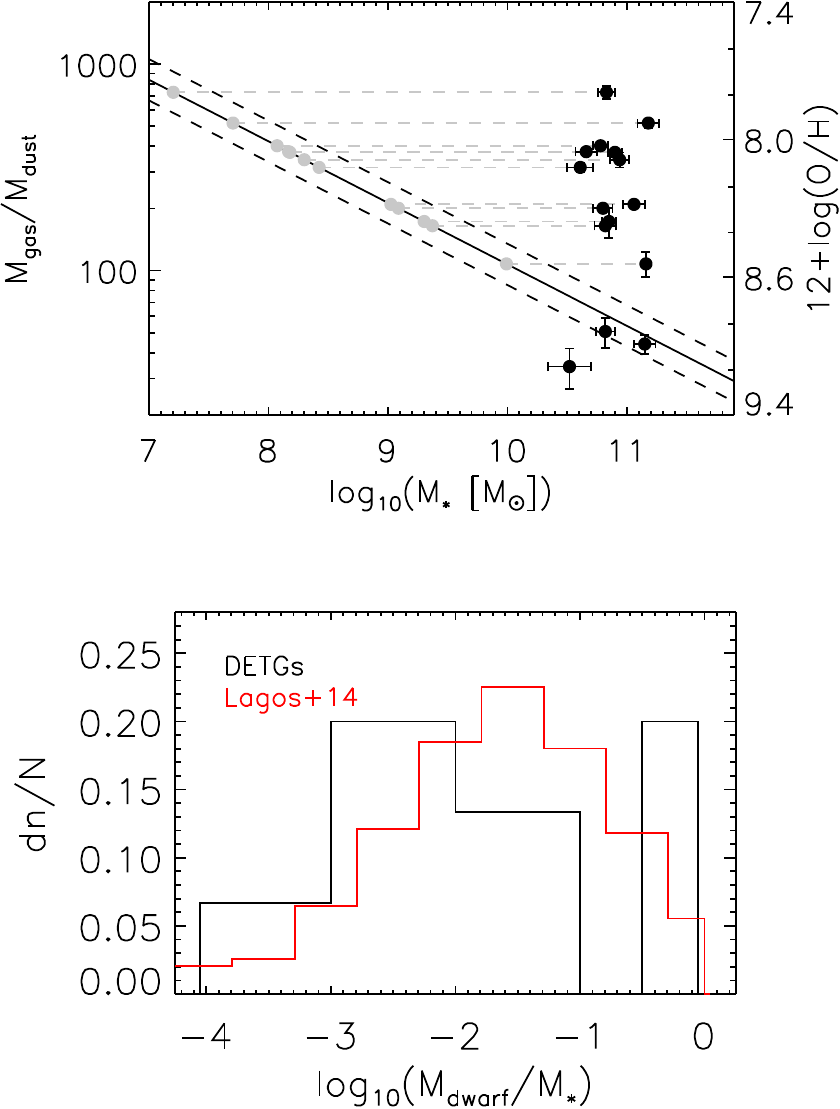}
 \end{center}
\caption{\textit{Top panel:} { The gas-to-dust ratio (plotted on the left axis) and the predicted gas-phase metallicity (right axis; see Section \ref{gasphasez})} of our sample DETGs plotted as a function of galaxy stellar mass (black points). Shown as a black line is the expected gas-to-dust ratio as a function of galaxy mass (using the mass metallicity relation for low mass and dwarf galaxies; \citealt{2006ApJ...647..970L}), with the 1$\sigma$ scatter from the mass-metallicity relation shown as dashed lines. Each point where the measured gas-to-dust ratio is higher than expected is connected to its expected minor merger partner mass by a grey dashed line. \textit{Bottom panel:} Histogram of the predicted merger mass ratios inferred as in the top panel. Also shown as a red histogram is the gas-rich merger mass fractions predicted from the SAMs of \protect \cite{2014MNRAS.443.1002L}. }
\label{g2dtoprog}
 \end{figure}

Further evidence for the external origin of this gas comes from the measured gas-to-dust ratios (listed in Table \ref{ratiotable}, and shown in Figure \ref{g2dfig}) and gas-phase metallicities (discussed in Section \ref{gasphasez}). Gas which has been produced through secular processes (e.g. stellar mass loss) would be expected to have a high metallicity, and thus a larger ratio of dust per unit gas mass than ISM acquired from mergers with smaller companions. Figure \ref{g2dfig} indicates the typical gas-to-dust ratio found in massive spiral galaxies, and early type galaxies from the \emph{Herschel} Reference Survey \citep{2012ApJ...748..123S}. Also shown is the measured gas-to-dust ratio in the Large Magellanic Cloud (LMC; \citealt{2010A&A...518L..74R}), which has a sub-solar metallicity. Our sources have a wide range of gas-to-dust ratios, suggesting the gas was not formed in situ in the majority of these DETGs.

 The top panel of Figure \ref{g2dtoprog} shows this in another way, plotting our DETG gas-to-dust ratios (and predicted gas-phase metallicities) as a function of stellar mass (black points). We also show the expected gas-to-dust ratio as a function of galaxy mass as a black line (assuming the gas-to-dust vs metallicity relation from Section \ref{gasphasez} and using the mass metallicity relation for low mass and dwarf galaxies; \citealt{2006ApJ...647..970L}). We choose to display the dwarf galaxy relation here, as we will utilise it in what follows. { Our extrapolation of this relation to high galaxy masses is not strictly valid, as above stellar masses of $\approx$10$^{10}$ \msun observed mass-metallicity relations are found to turn over (being flatter at high stellar masses; e.g. \citealt{2004ApJ...613..898T}), but this would not change the results we derive from this relation, as discussed below.}
Mass - metallicity relations are notably hard to calibrate, with large systematic offsets between different calibrations \citep[e.g.][]{2008ApJ...681.1183K}. As our galaxies have very similar masses, no matter which relation we take as input here, the spread in gas-to-dust ratio can only be explained by a varying metallicity.  
 
{ Three of our objects (DETG231, 920 and 967) lie on or below the black line in the top panel of Figure \ref{g2dtoprog}, having more dust per unit gas than expected even for their current galaxy mass. One objects has no \hi\ observations, and so its gas-to-dust ratio is likely to be higher, but the other two objects are genuinely dust rich. These galaxies have some of the most relaxed dust lanes in our sample (see Figure 1), and thus perhaps the gas is internally generated (and the tidal debris seen in deep imaging was from a gas poor minor merger). Alternatively the gas may have been brought in a sufficiently long time ago that signatures of a major merger have dissipated, or the metallicity of gas brought in through a minor merger has had time to increase. From this point on we assume that the gas in these objects was brought in through a equal mass major merger, but entirely removing these objects from our analysis does not change our conclusions.}  

If we assume, as discussed above, that the ETG present before the merger was completely devoid of gas, then all the gas and dust currently present must have been accreted from the minor merger. We can use the observed gas-to-dust ratio to infer the likely mass of the accreted smaller object, and the mass ratio of the merger.  Figure \ref{g2dtoprog} demonstrates how we do this. In the top panel the grey dashed lines connect the measured gas-to-dust ratio with the mass of the system that would be expected to have such a ratio. As many of the predicted stellar masses correspond to dwarf galaxy sized objects, we use the mass-metallicity relation for low-mass and dwarf galaxies from \citealt{2006ApJ...647..970L} to make these predictions. The bottom panel of Figure \ref{g2dtoprog} shows the predicted mass ratio of the merger as a black histogram. The median predicted mass ratio for the merger would be $\approx$40:1.

{ We note that the progenitor masses we derive have a high level of uncertainty. For instance, they are sensitive to our choice of mass-metallicity relation. If instead of using the result for dwarf galaxies (from \citealt{2006ApJ...647..970L}) we had used the relation more applicable to massive galaxies from \cite{2004ApJ...613..898T}, then our derived progenitor masses would be lower (by up to $\approx$0.8 dex at stellar masses around 10$^9$ \msun; less for smaller satellites). The mass-metallicity relation has also been found to have a secondary dependance on star-formation rate \citep[e.g.][]{2010A&A...521L..53L,2010MNRAS.408.2115M}. Our predicted progenitors are predicted to be gas rich (see below), and are thus likely to have been highly star-forming, and thus have a higher metallicity than expected for their stellar mass. This would lead us to underestimate the stellar mass of the progenitor (by upto $\approx$0.5 dex;  \citealt{2010MNRAS.408.2115M}).
Another uncertainty is the state of the ETG before the merger. If we relax the assumption that the parent ETG was gas poor, and instead allow it to have some (internally produced) gas and dust which has now mixed with the accreted gas, then the progenitor masses we derive would go down. Overall, we caution that the masses derived are indicative only, but go on (below) to find they do agree reasonably well with the predictions from simulations.}

In the bottom panel of Figure \ref{g2dtoprog} we show the predicted distribution of mass ratios for $z$=0 gas-rich mergers with ETGs, taken from the simulations of \cite{2014MNRAS.443.1002L}.
\cite{2014MNRAS.443.1002L} presented a detailed study of the atomic and molecular gas contents of model ETGs using a semi-analytic model (SAM) of galaxy formation in a $\Lambda$CDM universe. They extensively compared the properties of model ETGs with the results of ATLAS$^{\rm 3D}$. Their model reproduces well the fraction of ETGs as a function of stellar mass in the local Universe, as well as the \hi\ and H$_2$ content of those ETGs. This makes the Lagos et al. model a good tool to look into model ETGs that underwent a gas-rich merger.

In order to perform a fair comparison with our observations, we select ETGs in the model at $z=0$ (those with bulge-to-total stellar mass ratios$>$0.5) that have stellar masses $>$10$^{10}$ \msun\ and had a gas-rich (gas fraction of the merger system (M$_{\rm \hi}$+M$_{\rm H2}$)/M$_{\rm star} >$ 0.1) merger in the last $\approx$0.4 Gyr. If we now analyse the nature of this sample of ETGs, we see that as expected the models predict these gas-rich mergers to be essentially all minor mergers (with the mass ratio distribution shown in the bottom panel of \ref{g2dtoprog}). The other prediction is that most of the gas the model ETGs have after the minor merger was brought in by the satellite galaxy, typically increasing the total ISM mass of the coalesced system by a factor of $\approx$9, consistent with our assumption above.

Comparing the model with our data, we find that our estimated companion masses follow the expected trend, and a Mann-Whitney U-test is unable to reject the null-hypothesis that our proposed mergers are drawn from the same underlying mass ratio distribution as predicted from the SAMs (any differences are at less than a 1$\sigma$ level).
The mass ratio of the gas-rich mergers is a completely new constraint to the semi-analytic models, and can clearly reveal if the models may be having problems with the predicted gas contents of different galaxy populations. The fact that the Lagos et al. model described these observations well is encouraging as none of these observations were used to construct the model. The Lagos et al model, despite its simplifications, does seem to capture at least some of the behaviour seen in real galaxies.

\begin{figure}
\begin{center}
\includegraphics[width=0.45\textwidth,angle=0,clip,trim=0.0cm 0.0cm 0cm 0.0cm]{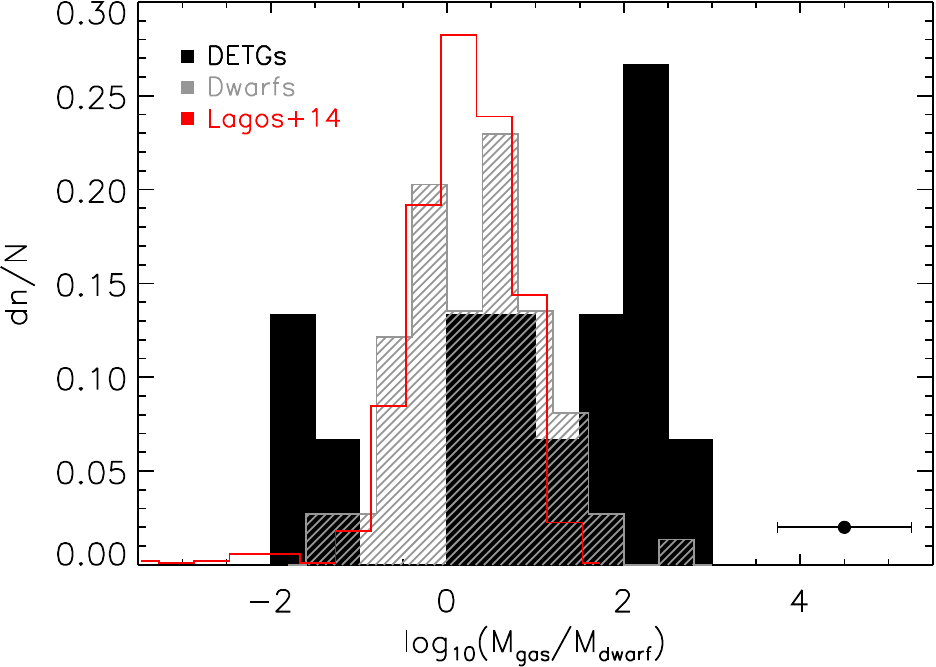}
 \end{center}
\caption{Histogram of the predicted gas fractions in the minor merger companion, assuming all the gas mass in our DETGs came from a minor merger with a dwarf companion (whose mass is derived as in Figure \ref{g2dtoprog}). Also shown as a shaded histogram are the observed gas fractions in dwarf galaxies from \protect \cite{2012AJ....143..133H}, and as a red histogram is the gas fractions predicted for gas rich mergers onto ETGs in the SAMs of \protect \cite{2014MNRAS.443.1002L}.}
\label{prog_gasfrac}
 \end{figure}

In Figure \ref{prog_gasfrac} we plot the gas fractions the gas-rich dwarf galaxy is predicted to have had before the merger as a black filled histogram.  The gas fractions of some of the accreted satellites are high, with the median accreted galaxy having $\approx$10 times more gas mass than stellar mass.  The error bars on this estimate are large, with a median error bar being displayed on the bottom right (we include the errors in gas and dust masses, and propagate these through, including a source of error from the scatter of the mass-metallicity relation). In Figure \ref{prog_gasfrac} we also plot the observed distribution of dwarf galaxy gas mass fractions from \cite{2012AJ....143..133H}, shown as a grey shaded histogram. We also show as a red histogram the predicted gas fractions of $z$=0 gas-rich mergers onto ETGs, again taken from the semi-analytic models (SAMs) of \cite{2014MNRAS.443.1002L}. The predicted and observed gas fractions (from \citealt{2012AJ....143..133H} and \citealt{2014MNRAS.443.1002L} respectively) are found to agree well with one another.

As Figure \ref{prog_gasfrac} shows, our predicted distribution of minor companion gas mass fractions are skewed to larger values than found in the sample of local dwarf galaxies. This can likely be explained by our strong selection function towards detecting more gas-rich objects, ensuring we are only selecting the most gas rich minor mergers in the local universe. It should be noted that the observed dwarf galaxy gas fractions actually only include atomic gas, thus the total gas mass is somewhat higher, decreasing any discrepancy. Dwarf galaxies as gas-rich as those required here do exist in the study of \cite{2012AJ....143..133H}, but are rare ($\approx$7\% of the observed dwarfs have gas to stellar mass fractions greater than 20). They also found $\approx$2\% of dwarfs had gas masses greater than 100 times their stellar mass.

\subsection{Suppression of star-formation in minor mergers}
\label{sfesection}
 \begin{figure}
\begin{center}
\includegraphics[width=0.45\textwidth,angle=0,clip,trim=0.0cm 0.0cm 0cm 0.0cm]{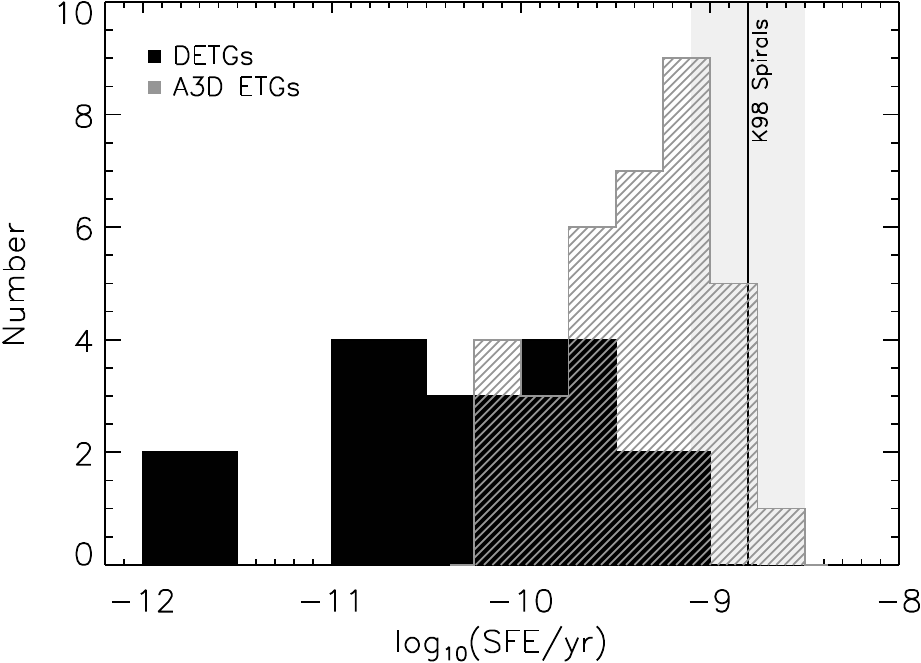}
 \end{center}
\caption{Histogram of the star formation efficiency (SFE=SFR/M$_{\rm H2}$) in our DETGs (black histogram) and \atlas\ ETGs (grey shaded histogram). The average SFE of the spiral galaxies from \protect \cite{1998ApJ...498..541K} is also shown as a black vertical line, with a grey bar showing the standard deviation of their SFEs around the mean.  }
\label{sfeplot}
 \end{figure}

In Table \ref{proptable} we presented the star-formation rates of the sample DETGs, estimated via SED fitting (using the \texttt{MAGPHYS} code). The derived values vary between 0.01 and 1.3~\msun\ yr$^{-1}$. These are reasonable star-formation rates for ETGs, well within the range found within the \atlas\ sample (0.008 - 18 \msun yr$^{-1}$; D14).

\citet[hereafter D14]{2014MNRAS.444.3427D} reported that the star-formation was suppressed in \atlas\ ETGs (compared to late-type galaxies), with their sample galaxies lying on average a factor of $\approx$2.5 below the Kennicutt-Schmidt relation (e.g. \citealt{1998ApJ...498..541K}). We are unable to place our galaxies on the Kennicutt-Schmidt relation, as we have no estimate of the size of the star-forming region in these objects. We can however estimate the star-formation efficiency (SFE; defined as star formation per unit gas mass), using our gas masses and derived star-formation rates. In what follows we define the SFE as star-formation rate per unit molecular gas mass (SFR/M$_{\rm H2}$) rather than total gas, because molecular gas is the phase which is most closely linked to star formation \citep[e.g.][]{2008AJ....136.2846B}. Including \hi\ (or using a variable X$_{\rm CO}$) in our SFEs would make the SFEs even smaller (or equivalently, make depletion times longer). 

Figure \ref{sfeplot} shows the SFE in our DETGs (black histogram), and the \atlas\ ETGs (grey histogram). Also shown (as a black line/grey bar) are the spiral galaxies of \cite{1998ApJ...498..541K}, which have been found to have average SFEs of $\approx$1.5$\times10^{-9}$ yr$^{-1}$ (see also e.g. \citealt{2011ApJ...730L..13B}). The distribution of SFE in our DETGs overlap with the distribution found in \atlas, but the medians are substantially different. Our DETG sample objects have median SFEs of $\approx$9$\times10^{-11}$  yr$^{-1}$, while the median for the entire sample of \atlas\ ETGs is closer to $\approx$4$\times10^{-10}$  yr$^{-1}$. A Mann-Whitney U test finds that these two SFE distributions are unlikely to be drawn from the same parent distribution at a confidence level $\gg$3$\sigma$. 
 Thus, if star formation were to continue on in a steady state at the same level as today, the gas reservoir in our objects would take a median time of $\approx$11 Gyr to be depleted. We note that the median SFE in our sample would be even lower if we use 
a metallicity-derived X$_{\rm CO}$ conversation factor. 

We checked that our results were not dependant on the method used to estimate star-formation rates by also calculating monochromatic star-formation rates from \textit{Wide-field Infrared Survey Explorer} (\textit{WISE}; \citealt{2010AJ....140.1868W}) 22~\mum\ emission alone, and the 22~\mum\ data combined with observations of far-ultraviolet emission from the \textit{Galaxy Evolution Explorer} (GALEX; \citealt{2007ApJS..173..682M}). We used the same procedure as described in \cite{2014MNRAS.444.3427D}, taking 22~\mum\ aperture magnitudes and foreground extinction corrected FUV magnitudes for the DETGs from the data compiled in K13. We note that the aperture values used include the entire galaxies, allowing us to properly subtract the circumstellar emission from old stars, which can be important at 22\mum\ (see \cite{2014MNRAS.444.3427D} for a full discussion of this process). These three estimates of the star-formation rate agree well, apart from at low SFRs where the 22\mum\ indicator gives slightly higher SFRs (at a 1$\sigma$ level). Whatever the method used to estimate the SFR, we find our DETGs have very low SFEs, with median depletion times in excess of 6.6 Gyr. 
 
This low star-formation efficiency in remnants of recent gas-rich minor mergers is puzzling. Major mergers are usually associated with strong starbursts at low redshift (e.g. Arp 220), with extreme star-formation rates, and fast depletion times (some as low as 10$^7$ yr; e.g. \citealt{2004ApJ...606..271G}). This is usually explained as a dissipative collapse of the gas in the galaxy to the centre, driving gas densities up, which then results in a nuclear starburst \citep[e.g.][]{Mihos:1996bo}. { The three objects in our sample which may have had major mergers (discussed in Section \ref{prog_estimate}) do lie at the higher end of our SFE distribution, but are not strong outliers}. Minor mergers are less violent events, but these too might be expected to increase the SFE by some factor (see e.g. \citealt{2012ApJ...758...73S} who report evidence for such a relation with mergers of all types). The DETGs studied here do not have strong AGN, supporting the idea that there is a large time delay (in merging systems) between the onset of star formation and AGN activity \citep{2012MNRAS.423...59S,2014arXiv1411.2028K}. We thus consider that AGN feedback is unlikely to be important in suppressing the star formation in these objects.

Studies such as \cite{2012ApJ...758...73S} and \cite{2014MNRAS.440.2944K} do find that the presence of a large bulge reduces the SFE (a conclusion supported by the low SFE of \atlas\ ETGs in \citealt{2014MNRAS.444.3427D}), and it has been posited that this effect may arise because the presence of a bulge keeps gas more stable against star formation (e.g. \citealt{2009ApJ...707..250M,2013MNRAS.432.1914M}). In our sample, the galaxies with a more `normal' SFE do tend to have strong disks, but so do some of the very suppressed objects. It is also unclear why this sample of recent minor merger remnants would show even greater SFE suppression than normal ETGs (both those from \atlas, and the massive galaxies in the \citealt{2012ApJ...758...73S} sample), as they likely have similar bulge masses. 
A relation between stellar mass and star-formation efficiency has been shown by \cite{2012ApJ...758...73S,2014MNRAS.445.2599B}, but our objects lie well off this relation. Stability enhancement by the bulge would also likely be most efficient on regular gas reservoirs which have been in place for some time. 

One possible explanation is that we have caught our objects in a special phase of their evolution, where gas is still streaming towards the centre (potentially being stabilised by these inflowing motions, e.g. \citealt{2013ApJ...779...45M}, or additional shocks/turbulence). The size of the star-forming region in this type of galaxy has been observed to decrease as the merger proceeds (Shabala et al., in prep.), supporting the idea the gas may be inflowing. Once gas densities in the centre of these objects increase a starburst may be triggered, which would lead us to no longer classify these galaxies as DETGs. Support for such a picture comes from observations of star-formation in tidal tails produced by minor mergers and ram-pressure stripping, where the SFE is also found to be low \citep[e.g.][]{2013ApJ...774..125K,2014ApJ...792...11J}. In addition, observations of interacting systems/compact groups suggest that if the ISM is kept warm by shocks and accretion induced turbulence, this may result in suppressed star-formation efficiencies \citep[e.g.][]{2014ApJ...795..159A, 2014ApJ...797..117A}.

Conversely, many of the CO line profiles shown in Figure \ref{codetsfig1} are (within our ability to measure them) symmetric and show double horns (that normally result from gas rotating regularly in the galaxy potential). Paper II presents more in depth analysis of the \hi\ line shape, which may be more sensitive to dynamical perturbations (because dynamical times are longer in large \hi\ disks). 
The dust disks visible in the optical images of these galaxies for the most part also seem regular. These two pieces of evidence argue against the interpretation above, where the ongoing merger is strongly dynamically perturbing the gas reservoir, but without resolved gas imaging it is hard to conclude what the correct physical explanation for the low SFE of our DETGs may be.

\section{Conclusions}
\label{conclude}

In this work we used the IRAM-30m telescope to detect CO emission in a sample of bulge-dominated galaxies which have large dust lanes visible in optical imaging, and far-infrared fluxes available from the \emph{Herschel} space telescope. These systems show disturbances in their optical light distribution that suggest they have had minor mergers in the recent past, which have supplied their gas and dust. 

We firmly detect 15 of our target objects in at least one CO line, and determine that these galaxies are very gas rich, having H$_2$ masses between 4$\times$10$^{8}$ and 2$\times$10$^{10}$ \msun. We use these molecular gas masses, combined with atomic gas masses (from Paper II in this series) to calculate gas-to-dust and gas-to-stellar mass ratios (using dust and stellar masses derived from SED fitting following K13).  We find molecular to stellar mass ratios (M$_{\rm H2}$ /M$_{\rm *}$) of between 0.8 and 23\%, and total gas to stellar mass ratios (M$_{\rm gas}$ /M$_{\rm *}$) between 1 and 51\%. 
The gas-to-dust ratios (M$_{\rm gas}$ /M$_{\rm d}$) of our sample objects vary widely, between values of $\approx$50 and 730 (see Figure \ref{g2dfig}). We discuss our assumption of a fixed X$_{\rm CO}$ for these objects given this variation, and conclude that we may be underestimating the molecular gas mass in many of our objects. Any such underestimation only serves to strengthen the results that follow.

The large range in gas-to-dust ratios in our objects implies a wide range of gas-phase metallicities, much larger than predicted by the scatter in the mass-metallicity relation. This suggests that the gas in these objects has indeed been accreted through a recent merger with a lower mass companion. We use the gas-to-dust to metallicity, and the mass-metallicity relations to calculate the implied minor companion masses and gas fractions, assuming the gas and dust have been accreted in their current proportions. We find the likely dwarf companions had masses as low as $\approx$10$^7$ \msun\ and that the implied merger mass ratios are consistent with the expectation from simulations. The median predicted mass ratio for the mergers is $\approx$40:1. The accreted satellites are predicted to be very gas rich, but are again consistent with being drawn from the population of known dwarf galaxies.

In Section \ref{sfesection} we showed that (no matter which SFR indicator is used) our sample objects have very low star-formation efficiencies, taking on average 6.5 Gyr to deplete their gas reservoirs. This cannot be purely a result of morphological quenching mechanisms, as this level of suppression is greater than found in the \atlas\ sample of ETGs. We discuss mechanisms that could cause such a suppression, include dynamical effects caused by the minor merger, but with the current data available are unable to reach a firm conclusion.

These results clearly motivate further study of the minor merger process, and its effect on star formation in the remnant. Resolved observations will be required to understand the configuration of the gas in these objects and enable resolved studies of their star-formation efficiency. Simulations of gas-rich minor companions merging with gas-poor ETGs will help determine if the observed suppression of star formation is caused by the merger itself, and thus if it is a more general process that has importance across the Hubble sequence.

\noindent \textbf{Acknowledgments}

TAD acknowledges support from a Science and Technology Facilities Council Ernest Rutherford Fellowship, and thanks Maarten Baes, Gianfranco De Zotti, Iv\'an Oteo G\'omez, Michal Michalowski and Catherine Vlahakis for comments which improved the paper. KR acknowledges support from the European Research Council Starting
 Grant SEDmorph (P.I. V.~Wild). SSS thanks the Australian Research Council for an Early Career Fellowship (DE130101399). LD, RJI and SJM acknowledge support from the European Research Council Advanced grant COSMICISM.
The research leading to these results has received funding from the European
Community's Seventh Framework Programme (/FP7/2007-2013/) under grant agreement
No 229517 and No 283393 (RadioNet3).
This paper is based on observations carried out with the IRAM Thirty Meter Telescope. IRAM is supported by INSU/CNRS (France), MPG (Germany) and IGN (Spain). 

The \emph{Herschel}-ATLAS is a project with \emph{Herschel}, which is an ESA space
observatory with science instruments provided by European-led
Principal Investigator consortia and with important participation from
NASA. The H-ATLAS website is http://www.h-atlas.org/. he GAMA input
catalogue is based on data taken from the Sloan Digital Sky Survey and
the UKIRT Infrared Deep Sky Survey. Complementary imaging of the GAMA
regions is being obtained by a number of independent survey programs
including \emph{GALEX} MIS, VST KIDS, VISTA VIKING, \emph{WISE}, \emph{Herschel}-ATLAS,
GMRT and ASKAP providing UV to radio coverage. GAMA is funded by the
STFC (UK), the ARC (Australia), the AAO, and the participating
institutions. The GAMA website is http://www.gama-survey.org/.
This publication makes use of data products from the Wide-field Infrared Survey Explorer, which is a joint project of the University of California, Los Angeles, and the Jet Propulsion Laboratory/California Institute of Technology, funded by the National Aeronautics and Space Administration.
This research has made use of the NASA/IPAC Extragalactic Database (NED) which is operated by the Jet Propulsion Laboratory, California Institute of Technology, under contract with the National Aeronautics and Space Administration.

\bsp
\bibliographystyle{mn2e}
\bibliography{1DETGbib.bib}

\begin{thebibliography}{92}
\expandafter\ifx\csname natexlab\endcsname\relax\def\natexlab#1{#1}\fi

\bibitem[{Alatalo {et~al}\mbox{.}(2014)Alatalo, Appleton, Lisenfeld, Bitsakis,
  Guillard, Charmandaris, Cluver, Dopita, Freeland, Jarrett, Kewley, Ogle,
  Rasmussen, Rich, Verdes-Montenegro, Xu, \& Yun}]{2014ApJ...795..159A}
Alatalo K. {et~al.}, 2014, ApJ, 795, 159

\bibitem[{Appleton {et~al}\mbox{.}(2014)Appleton, Mundell, Bitsakis, Lacy,
  Alatalo, Armus, Charmandaris, Duc, Lisenfeld, \& Ogle}]{2014ApJ...797..117A}
Appleton P.~N. {et~al.}, 2014, ApJ, 797, 117

\bibitem[{Asplund {et~al}\mbox{.}(2004)Asplund, Grevesse, Sauval,
  Allende~Prieto, \& Kiselman}]{2004A&A...417..751A}
Asplund M., Grevesse N., Sauval A.~J., Allende~Prieto C., Kiselman D., 2004,
  A\&A, 417, 751

\bibitem[{Bigiel {et~al}\mbox{.}(2008)Bigiel, Leroy, Walter, Brinks, de~Blok,
  Madore, \& Thornley}]{2008AJ....136.2846B}
Bigiel F., Leroy A., Walter F., Brinks E., de~Blok W. J.~G., Madore B.,
  Thornley M.~D., 2008, AJ, 136, 2846

\bibitem[{Bigiel {et~al}\mbox{.}(2011)Bigiel, Leroy, Walter, Brinks, de~Blok,
  Kramer, Rix, Schruba, Schuster, Usero, \& Wiesemeyer}]{2011ApJ...730L..13B}
Bigiel F. {et~al.}, 2011, ApJL, 730, L13

\bibitem[{Bolatto {et~al}\mbox{.}(2008)Bolatto, Leroy, Rosolowsky, Walter, \&
  Blitz}]{2008ApJ...686..948B}
Bolatto A.~D., Leroy A.~K., Rosolowsky E., Walter F., Blitz L., 2008, ApJ, 686,
  948

\bibitem[{Bolatto, Wolfire \& Leroy(2013)Bolatto, Wolfire, \&
  Leroy}]{2013ARA&A..51..207B}
Bolatto A.~D., Wolfire M., Leroy A.~K., 2013, ARAA, 51, 207

\bibitem[{Boselli {et~al}\mbox{.}(2014)Boselli, Cortese, Boquien, Boissier,
  Catinella, Lagos, \& Saintonge}]{2014A&A...564A..66B}
Boselli A., Cortese L., Boquien M., Boissier S., Catinella B., Lagos C.,
  Saintonge A., 2014, A\&A, 564, A66

\bibitem[{Bothwell {et~al}\mbox{.}(2014)Bothwell, Wagg, Cicone, Maiolino,
  M{\o}ller, Aravena, De~Breuck, Peng, Espada, Hodge, Impellizzeri,
  Mart{\'\i}n, Riechers, \& Walter}]{2014MNRAS.445.2599B}
Bothwell M.~S. {et~al.}, 2014, MNRAS, 445, 2599

\bibitem[{Braine \& Combes(1992)}]{1992A&A...264..433B}
Braine J., Combes F., 1992, A\&A, 264, 433

\bibitem[{Bruzual \& Charlot(2003)}]{2003MNRAS.344.1000B}
Bruzual G., Charlot S., 2003, MNRAS, 344, 1000

\bibitem[{Cappellari {et~al}\mbox{.}(2011)Cappellari, Emsellem, Krajnovic,
  McDermid, Scott, Verdoes~Kleijn, Young, Alatalo, Bacon, Blitz, Bois,
  Bournaud, Bureau, Davies, Davis, de~Zeeuw, Duc, Khochfar, Kuntschner,
  Lablanche, Morganti, Naab, Oosterloo, Sarzi, Serra, \&
  Weijmans}]{2011MNRAS.413..813C}
Cappellari M. {et~al.}, 2011, MNRAS, 413, 813

\bibitem[{Chabrier(2003)}]{2003PASP..115..763C}
Chabrier G., 2003, PASP, 115, 763

\bibitem[{Combes, Young \& Bureau(2007)Combes, Young, \&
  Bureau}]{2007MNRAS.377.1795C}
Combes F., Young L.~M., Bureau M., 2007, MNRAS, 377, 1795

\bibitem[{Crocker {et~al}\mbox{.}(2012)Crocker, Krips, Bureau, Young, Davis,
  Bayet, Alatalo, Blitz, Bois, Bournaud, Cappellari, Davies, de~Zeeuw, Duc,
  Emsellem, Khochfar, Krajnovic, Kuntschner, Lablanche, McDermid, Morganti,
  Naab, Oosterloo, Sarzi, Scott, Serra, \& Weijmans}]{2012MNRAS.421.1298C}
Crocker A. {et~al.}, 2012, MNRAS, 421, 1298

\bibitem[{Crocker {et~al}\mbox{.}(2011)Crocker, Bureau, Young, \&
  Combes}]{Crocker:2011ic}
Crocker A.~F., Bureau M., Young L.~M., Combes F., 2011, MNRAS, 410, 1197

\bibitem[{Da~Cunha, Charlot \& Elbaz(2008)Da~Cunha, Charlot, \&
  Elbaz}]{2008MNRAS.388.1595D}
Da~Cunha E., Charlot S., Elbaz D., 2008, MNRAS, 388, 1595

\bibitem[{Davis {et~al}\mbox{.}(2013)Davis, Alatalo, Bureau, Cappellari, Scott,
  Young, Blitz, Crocker, Bayet, Bois, Bournaud, Davies, de~Zeeuw, Duc,
  Emsellem, Khochfar, Krajnovic, Kuntschner, Lablanche, McDermid, Morganti,
  Naab, Oosterloo, Sarzi, Serra, \& Weijmans}]{2013MNRAS.429..534D}
Davis T.~A. {et~al.}, 2013, MNRAS, 429, 534

\bibitem[{Davis {et~al}\mbox{.}(2011{\natexlab{a}})Davis, Alatalo, Sarzi,
  Bureau, Young, Blitz, Serra, Crocker, Krajnovic, McDermid, Bois, Bournaud,
  Cappellari, Davies, Duc, de~Zeeuw, Emsellem, Khochfar, Kuntschner, Lablanche,
  Morganti, Naab, Oosterloo, Scott, \& Weijmans}]{2011MNRAS.417..882D}
Davis T.~A. {et~al.}, 2011{\natexlab{a}}, MNRAS, 417, 882

\bibitem[{Davis {et~al}\mbox{.}(2011{\natexlab{b}})Davis, Bureau, Young,
  Alatalo, Blitz, Cappellari, Scott, Bois, Bournaud, Davies, de~Zeeuw,
  Emsellem, Khochfar, Krajnovic, Kuntschner, Lablanche, McDermid, Morganti,
  Naab, Oosterloo, Sarzi, Serra, \& Weijmans}]{Davis:2011bg}
Davis T.~A. {et~al.}, 2011{\natexlab{b}}, MNRAS, 414, 968

\bibitem[{Davis {et~al}\mbox{.}(2014)Davis, Young, Crocker, Bureau, Blitz,
  Alatalo, Emsellem, Naab, Bayet, Bois, Bournaud, Cappellari, Davies, de~Zeeuw,
  Duc, Khochfar, Krajnovic, Kuntschner, McDermid, Morganti, Oosterloo, Sarzi,
  Scott, Serra, \& Weijmans}]{2014MNRAS.444.3427D}
Davis T.~A. {et~al.}, 2014, MNRAS, 444, 3427

\bibitem[{Dickman, Snell \& Schloerb(1986)Dickman, Snell, \&
  Schloerb}]{Dickman:1986jz}
Dickman R.~L., Snell R.~L., Schloerb F.~P., 1986, ApJ, 309, 326

\bibitem[{Downes \& Solomon(1998)}]{1998ApJ...507..615D}
Downes D., Solomon P.~M., 1998, ApJ, 507, 615

\bibitem[{Draine {et~al}\mbox{.}(2007)Draine, Dale, Bendo, Gordon, Smith,
  Armus, Engelbracht, Helou, Kennicutt, Li, Roussel, Walter, Calzetti,
  Moustakas, Murphy, Rieke, Bot, Hollenbach, Sheth, \&
  Teplitz}]{2007ApJ...663..866D}
Draine B.~T. {et~al.}, 2007, ApJ, 663, 866

\bibitem[{Driver {et~al}\mbox{.}(2011)Driver, Hill, Kelvin, Robotham, Liske,
  Norberg, Baldry, Bamford, Hopkins, Loveday, Peacock, Andrae, Bland-Hawthorn,
  Brough, Brown, Cameron, Ching, Colless, Conselice, Croom, Cross, de~Propris,
  Dye, Drinkwater, Ellis, Graham, Grootes, Gunawardhana, Jones, van Kampen,
  Maraston, Nichol, Parkinson, Phillipps, Pimbblet, Popescu, Prescott,
  Roseboom, Sadler, Sansom, Sharp, Smith, Taylor, Thomas, Tuffs, Wijesinghe,
  Dunne, Frenk, Jarvis, Madore, Meyer, Seibert, Staveley-Smith, Sutherland, \&
  Warren}]{2011MNRAS.413..971D}
Driver S.~P. {et~al.}, 2011, MNRAS, 413, 971

\bibitem[{Duc {et~al}\mbox{.}(2015)Duc, Cuillandre, Karabal, Cappellari,
  Alatalo, Blitz, Bournaud, Bureau, Crocker, Davies, Davis, de~Zeeuw, Emsellem,
  Khochfar, Krajnovic, Kuntschner, McDermid, Michel-Dansac, Morganti, Naab,
  Oosterloo, Paudel, Sarzi, Scott, Serra, Weijmans, \&
  Young}]{2015MNRAS.446..120D}
Duc P.-A. {et~al.}, 2015, MNRAS, 446, 120

\bibitem[{Eales {et~al}\mbox{.}(2010)Eales, Dunne, Clements, Cooray, De~Zotti,
  Dye, Ivison, Jarvis, Lagache, Maddox, Negrello, Serjeant, Thompson, van
  Kampen, Amblard, Andreani, Baes, Beelen, Bendo, Benford, Bertoldi, Bock,
  Bonfield, Boselli, Bridge, Buat, Burgarella, Carlberg, Cava, Chanial,
  Charlot, Christopher, Coles, Cortese, Dariush, da~Cunha, Dalton, Danese,
  Dannerbauer, Driver, Dunlop, Fan, Farrah, Frayer, Frenk, Geach, Gardner,
  Gomez, Gonz{\'a}lez-Nuevo, Gonz{\'a}lez-Solares, Griffin, Hardcastle,
  Hatziminaoglou, Herranz, Hughes, Ibar, Jeong, Lacey, Lapi, Lawrence, Lee,
  Leeuw, Liske, L{\'o}pez-Caniego, M{\"u}ller, Nandra, Panuzzo, Papageorgiou,
  Patanchon, Peacock, Pearson, Phillipps, Pohlen, Popescu, Rawlings, Rigby,
  Rigopoulou, Robotham, Rodighiero, Sansom, Schulz, Scott, Smith, Sibthorpe,
  Smail, Stevens, Sutherland, Takeuchi, Tedds, Temi, Tuffs, Trichas, Vaccari,
  Valtchanov, van~der Werf, Verma, Vieria, Vlahakis, \&
  White}]{2010PASP..122..499E}
Eales S. {et~al.}, 2010, PASP, 122, 499

\bibitem[{Gao \& Solomon(2004)}]{2004ApJ...606..271G}
Gao Y., Solomon P.~M., 2004, ApJ, 606, 271

\bibitem[{Haynes {et~al}\mbox{.}(2011)Haynes, Giovanelli, Martin, Hess,
  Saintonge, Adams, Hallenbeck, Hoffman, Huang, Kent, Koopmann, Papastergis,
  Stierwalt, Balonek, Craig, Higdon, Kornreich, Miller, O'Donoghue, Olowin,
  Rosenberg, Spekkens, Troischt, \& Wilcots}]{2011AJ....142..170H}
Haynes M.~P. {et~al.}, 2011, AJ, 142, 170

\bibitem[{Huang {et~al}\mbox{.}(2012)Huang, Haynes, Giovanelli, Brinchmann,
  Stierwalt, \& Neff}]{2012AJ....143..133H}
Huang S., Haynes M.~P., Giovanelli R., Brinchmann J., Stierwalt S., Neff S.~G.,
  2012, AJ, 143, 133

\bibitem[{Ibar {et~al}\mbox{.}(2010)Ibar, Ivison, Cava, Rodighiero,
  Buttiglione, Temi, Frayer, Fritz, Leeuw, Baes, Rigby, Verma, Serjeant,
  M{\"u}ller, Auld, Dariush, Dunne, Eales, Maddox, Panuzzo, Pascale, Pohlen,
  Smith, De~Zotti, Vaccari, Hopwood, Cooray, Burgarella, \&
  Jarvis}]{2010MNRAS.409...38I}
Ibar E. {et~al.}, 2010, MNRAS, 409, 38

\bibitem[{Israel(1998)}]{1998A&ARv...8..237I}
Israel F.~P., 1998, ARAA, 8, 237

\bibitem[{J{\'a}chym {et~al}\mbox{.}(2014)J{\'a}chym, Combes, Cortese, Sun, \&
  Kenney}]{2014ApJ...792...11J}
J{\'a}chym P., Combes F., Cortese L., Sun M., Kenney J. D.~P., 2014, ApJ, 792,
  11

\bibitem[{Katkov, Sil'chenko \& Afanasiev(2014)Katkov, Sil'chenko, \&
  Afanasiev}]{2014MNRAS.438.2798K}
Katkov I.~Y., Sil'chenko O.~K., Afanasiev V.~L., 2014, MNRAS, 438, 2798

\bibitem[{Kaviraj(2014{\natexlab{a}})}]{2014MNRAS.440.2944K}
Kaviraj S., 2014{\natexlab{a}}, MNRAS, 440, 2944

\bibitem[{Kaviraj(2014{\natexlab{b}})}]{2014MNRAS.437L..41K}
Kaviraj S., 2014{\natexlab{b}}, MNRASL, 437, L41

\bibitem[{Kaviraj {et~al}\mbox{.}(2009)Kaviraj, Peirani, Khochfar, Silk, \&
  Kay}]{2009MNRAS.394.1713K}
Kaviraj S., Peirani S., Khochfar S., Silk J., Kay S., 2009, MNRAS, 394, 1713

\bibitem[{Kaviraj {et~al}\mbox{.}(2013)Kaviraj, Rowlands, Alpaslan, Dunne,
  Ting, Bureau, Shabala, Lintott, Smith, Agius, Auld, Baes, Bourne, Cava,
  Clements, Cooray, Dariush, De~Zotti, Driver, Eales, Hopwood, Hoyos, Ibar,
  Maddox, Micha{\l}owski, Sansom, Smith, \& Valiante}]{2013MNRAS.435.1463K}
Kaviraj S. {et~al.}, 2013, MNRAS, 435, 1463

\bibitem[{Kaviraj {et~al}\mbox{.}(2007)Kaviraj, Schawinski, Devriendt,
  Ferreras, Khochfar, Yoon, Yi, Deharveng, Boselli, Barlow, Conrow, Forster,
  Friedman, \& Martin}]{2007ApJS..173..619K}
Kaviraj S. {et~al.}, 2007, ApJS, 173, 619

\bibitem[{Kaviraj {et~al}\mbox{.}(2014)Kaviraj, Shabala, Deller, \&
  Middelberg}]{2014arXiv1411.2028K}
Kaviraj S., Shabala S.~S., Deller A.~T., Middelberg E., 2014, arXiv, 2028

\bibitem[{Kaviraj {et~al}\mbox{.}(2011)Kaviraj, Tan, Ellis, \&
  Silk}]{2011MNRAS.411.2148K}
Kaviraj S., Tan K.-M., Ellis R.~S., Silk J., 2011, MNRAS, 411, 2148

\bibitem[{Kaviraj {et~al}\mbox{.}(2012)Kaviraj, Ting, Bureau, Shabala,
  Crockett, Silk, Lintott, Smith, Keel, Masters, Schawinski, \&
  Bamford}]{2012MNRAS.423...49K}
Kaviraj S. {et~al.}, 2012, MNRAS, 423, 49

\bibitem[{Kennicutt(1998)}]{1998ApJ...498..541K}
Kennicutt R.~C., 1998, ApJ, 498, 541

\bibitem[{Kewley \& Ellison(2008)}]{2008ApJ...681.1183K}
Kewley L.~J., Ellison S.~L., 2008, ApJ, 681, 1183

\bibitem[{Knapp {et~al}\mbox{.}(1989)Knapp, Guhathakurta, Kim, \&
  Jura}]{1989ApJS...70..329K}
Knapp G.~R., Guhathakurta P., Kim D.-W., Jura M.~A., 1989, ApJS, 70, 329

\bibitem[{Knierman {et~al}\mbox{.}(2013)Knierman, Scowen, Veach, Groppi,
  Mullan, Konstantopoulos, Knezek, \& Charlton}]{2013ApJ...774..125K}
Knierman K.~A., Scowen P., Veach T., Groppi C., Mullan B., Konstantopoulos I.,
  Knezek P.~M., Charlton J., 2013, ApJ, 774, 125

\bibitem[{Lagos {et~al}\mbox{.}(2014)Lagos, Davis, Lacey, Zwaan, Baugh,
  Gonzalez-Perez, \& Padilla}]{2014MNRAS.443.1002L}
Lagos C. d.~P., Davis T.~A., Lacey C.~G., Zwaan M.~A., Baugh C.~M.,
  Gonzalez-Perez V., Padilla N.~D., 2014, MNRAS, 443, 1002

\bibitem[{Lagos {et~al}\mbox{.}(2015)Lagos, Padilla, Davis, Lacey, Baugh,
  Gonzalez-Perez, Zwaan, \& Contreras}]{2015MNRAS.448.1271L}
Lagos C. d.~P., Padilla N.~D., Davis T.~A., Lacey C.~G., Baugh C.~M.,
  Gonzalez-Perez V., Zwaan M.~A., Contreras S., 2015, MNRAS, 448, 1271

\bibitem[{Lara-L{\'o}pez {et~al}\mbox{.}(2010)Lara-L{\'o}pez, Cepa,
  Bongiovanni, P{\'e}rez~Garc{\'\i}a, Ederoclite, Casta{\~n}eda,
  Fern{\'a}ndez~Lorenzo, Povi{\'c}, \&
  S{\'a}nchez-Portal}]{2010A&A...521L..53L}
Lara-L{\'o}pez M.~A. {et~al.}, 2010, A\&A, 521, L53

\bibitem[{Lee {et~al}\mbox{.}(2006)Lee, Skillman, Cannon, Jackson, Gehrz,
  Polomski, \& Woodward}]{2006ApJ...647..970L}
Lee H., Skillman E.~D., Cannon J.~M., Jackson D.~C., Gehrz R.~D., Polomski
  E.~F., Woodward C.~E., 2006, ApJ, 647, 970

\bibitem[{Leroy {et~al}\mbox{.}(2007)Leroy, Bolatto, Stanimirovic, Mizuno,
  Israel, \& Bot}]{2007ApJ...658.1027L}
Leroy A., Bolatto A., Stanimirovic S., Mizuno N., Israel F., Bot C., 2007, ApJ,
  658, 1027

\bibitem[{Leroy {et~al}\mbox{.}(2011)Leroy, Bolatto, Gordon, Sandstrom,
  Gratier, Rosolowsky, Engelbracht, Mizuno, Corbelli, Fukui, \&
  Kawamura}]{2011ApJ...737...12L}
Leroy A.~K. {et~al.}, 2011, ApJ, 737, 12

\bibitem[{Lintott {et~al}\mbox{.}(2008)Lintott, Schawinski, Slosar, Land,
  Bamford, Thomas, Raddick, Nichol, Szalay, Andreescu, Murray, \&
  Vandenberg}]{2008MNRAS.389.1179L}
Lintott C.~J. {et~al.}, 2008, MNRAS, 389, 1179

\bibitem[{Mannucci {et~al}\mbox{.}(2010)Mannucci, Cresci, Maiolino, Marconi, \&
  Gnerucci}]{2010MNRAS.408.2115M}
Mannucci F., Cresci G., Maiolino R., Marconi A., Gnerucci A., 2010, MNRAS, 408,
  2115

\bibitem[{Martig {et~al}\mbox{.}(2009)Martig, Bournaud, Teyssier, \&
  Dekel}]{2009ApJ...707..250M}
Martig M., Bournaud F., Teyssier R., Dekel A., 2009, ApJ, 707, 250

\bibitem[{Martig {et~al}\mbox{.}(2013)Martig, Crocker, Bournaud, Emsellem,
  Gabor, Alatalo, Blitz, Bois, Bureau, Cappellari, Davies, Davis, Dekel,
  de~Zeeuw, Duc, Falcon-Barroso, Khochfar, Krajnovic, Kuntschner, Morganti,
  McDermid, Naab, Oosterloo, Sarzi, Scott, Serra, Griffin, Teyssier, Weijmans,
  \& Young}]{2013MNRAS.432.1914M}
Martig M. {et~al.}, 2013, MNRAS, 432, 1914

\bibitem[{McDermid {et~al}\mbox{.}(2006)McDermid, Emsellem, Shapiro, Bacon,
  Bureau, Cappellari, Davies, de~Zeeuw, Falc{\'o}n-Barroso, Krajnovi{\'c},
  Kuntschner, Peletier, \& Sarzi}]{McDermid:2006bj}
McDermid R.~M. {et~al.}, 2006, MNRAS, 373, 906

\bibitem[{Meidt {et~al}\mbox{.}(2013)Meidt, Schinnerer, Garcia-Burillo, Hughes,
  Colombo, Pety, Dobbs, Schuster, Kramer, Leroy, Dumas, \&
  Thompson}]{2013ApJ...779...45M}
Meidt S.~E. {et~al.}, 2013, ApJ, 779, 45

\bibitem[{Merluzzi(1998)}]{1998A&A...338..807M}
Merluzzi P., 1998, A\&A, 338, 807

\bibitem[{Mihos \& Hernquist(1996)}]{Mihos:1996bo}
Mihos J.~C., Hernquist L., 1996, ApJ, 464, 641

\bibitem[{Morganti {et~al}\mbox{.}(2006)Morganti, de~Zeeuw, Oosterloo,
  McDermid, Krajnovi{\'c}, Cappellari, Kenn, Weijmans, \&
  Sarzi}]{2006MNRAS.371..157M}
Morganti R. {et~al.}, 2006, MNRAS, 371, 157

\bibitem[{Morrissey {et~al}\mbox{.}(2007)Morrissey, Conrow, Barlow, Small,
  Seibert, Wyder, Budav{\'a}ri, Arnouts, Friedman, Forster, Martin, Neff,
  Schiminovich, Bianchi, Donas, Heckman, Lee, Madore, Milliard, Rich, Szalay,
  Welsh, \& Yi}]{2007ApJS..173..682M}
Morrissey P. {et~al.}, 2007, ApJS, 173, 682

\bibitem[{Oosterloo {et~al}\mbox{.}(2002)Oosterloo, Morganti, Sadler, Vergani,
  \& Caldwell}]{2002AJ....123..729O}
Oosterloo T.~A., Morganti R., Sadler E.~M., Vergani D., Caldwell N., 2002, AJ,
  123, 729

\bibitem[{Parkin {et~al}\mbox{.}(2012)Parkin, Wilson, Foyle, Baes, Bendo,
  Boselli, Boquien, Cooray, Cormier, Davies, Eales, Galametz, Gomez,
  Lebouteiller, Madden, Mentuch, Page, Pohlen, Remy, Roussel, Sauvage, Smith,
  \& Spinoglio}]{2012MNRAS.422.2291P}
Parkin T.~J. {et~al.}, 2012, MNRAS, 422, 2291

\bibitem[{Pascale {et~al}\mbox{.}(2011)Pascale, Auld, Dariush, Dunne, Eales,
  Maddox, Panuzzo, Pohlen, Smith, Buttiglione, Cava, Clements, Cooray, Dye,
  De~Zotti, Fritz, Hopwood, Ibar, Ivison, Jarvis, Leeuw, L{\'o}pez-Caniego,
  Rigby, Rodighiero, Scott, Smith, Temi, Vaccari, \&
  Valtchanov}]{2011MNRAS.415..911P}
Pascale E. {et~al.}, 2011, MNRAS, 415, 911

\bibitem[{Pilbratt {et~al}\mbox{.}(2010)Pilbratt, Riedinger, Passvogel, Crone,
  Doyle, Gageur, Heras, Jewell, Metcalfe, Ott, \&
  Schmidt}]{2010A&A...518L...1P}
Pilbratt G.~L. {et~al.}, 2010, A\&A, 518, L1

\bibitem[{Rigby {et~al}\mbox{.}(2011)Rigby, Maddox, Dunne, Negrello, Smith,
  Gonz{\'a}lez-Nuevo, Herranz, L{\'o}pez-Caniego, Auld, Buttiglione, Baes,
  Cava, Cooray, Clements, Dariush, De~Zotti, Dye, Eales, Frayer, Fritz,
  Hopwood, Ibar, Ivison, Jarvis, Panuzzo, Pascale, Pohlen, Rodighiero,
  Serjeant, Temi, \& Thompson}]{2011MNRAS.415.2336R}
Rigby E.~E. {et~al.}, 2011, MNRAS, 415, 2336

\bibitem[{Roberts {et~al}\mbox{.}(1991)Roberts, Hogg, Bregman, Forman, \&
  Jones}]{1991ApJS...75..751R}
Roberts M.~S., Hogg D.~E., Bregman J.~N., Forman W.~R., Jones C., 1991, ApJS,
  75, 751

\bibitem[{Roman-Duval {et~al}\mbox{.}(2010)Roman-Duval, Israel, Bolatto,
  Hughes, Leroy, Meixner, Gordon, Madden, Paradis, Kawamura, Li, Sauvage, Wong,
  Bernard, Engelbracht, Hony, Kim, Misselt, Okumura, Ott, Panuzzo, Pineda,
  Reach, \& Rubio}]{2010A&A...518L..74R}
Roman-Duval J. {et~al.}, 2010, A\&A, 518, L74

\bibitem[{Rowlands {et~al}\mbox{.}(2012)Rowlands, Dunne, Maddox, Bourne, Gomez,
  Kaviraj, Bamford, Brough, Charlot, da~Cunha, Driver, Eales, Hopkins, Kelvin,
  Nichol, Sansom, Sharp, Smith, Temi, van~der Werf, Baes, Cava, Cooray, Croom,
  Dariush, De~Zotti, Dye, Fritz, Hopwood, Ibar, Ivison, Liske, Loveday, Madore,
  Norberg, Popescu, Rigby, Robotham, Rodighiero, Seibert, \&
  Tuffs}]{2012MNRAS.419.2545R}
Rowlands K. {et~al.}, 2012, MNRAS, 419, 2545

\bibitem[{Sage \& Welch(2006)}]{Sage:2006ko}
Sage L.~J., Welch G.~A., 2006, ApJ, 644, 850

\bibitem[{Saintonge {et~al}\mbox{.}(2012)Saintonge, Tacconi, Fabello, Wang,
  Catinella, Genzel, Graci{\'a}-Carpio, Kramer, Moran, Heckman, Schiminovich,
  Schuster, \& Wuyts}]{2012ApJ...758...73S}
Saintonge A. {et~al.}, 2012, ApJ, 758, 73

\bibitem[{Salim \& Rich(2010)}]{2010ApJ...714L.290S}
Salim S., Rich R.~M., 2010, ApJL, 714, L290

\bibitem[{Sandstrom {et~al}\mbox{.}(2013)Sandstrom, Leroy, Walter, Bolatto,
  Croxall, Draine, Wilson, Wolfire, Calzetti, Kennicutt, Aniano, Donovan~Meyer,
  Usero, Bigiel, Brinks, de~Blok, Crocker, Dale, Engelbracht, Galametz, Groves,
  Hunt, Koda, Kreckel, Linz, Meidt, Pellegrini, Rix, Roussel, Schinnerer,
  Schruba, Schuster, Skibba, van~der Laan, Appleton, Armus, Brandl, Gordon,
  Hinz, Krause, Montiel, Sauvage, Schmiedeke, Smith, \&
  Vigroux}]{2013ApJ...777....5S}
Sandstrom K.~M. {et~al.}, 2013, ApJ, 777, 5

\bibitem[{Sarzi {et~al}\mbox{.}(2006)Sarzi, Falc{\'o}n-Barroso, Davies, Bacon,
  Bureau, Cappellari, de~Zeeuw, Emsellem, Fathi, Krajnovi{\'c}, Kuntschner,
  McDermid, \& Peletier}]{Sarzi:2006p1474}
Sarzi M. {et~al.}, 2006, MNRAS, 366, 1151

\bibitem[{Serra {et~al}\mbox{.}(2012)Serra, Oosterloo, Morganti, Alatalo,
  Blitz, Bois, Bournaud, Bureau, Cappellari, Crocker, Davies, Davis, de~Zeeuw,
  Duc, Emsellem, Khochfar, Krajnovic, Kuntschner, Lablanche, McDermid, Naab,
  Sarzi, Scott, Trager, Weijmans, \& Young}]{2012MNRAS.422.1835S}
Serra P. {et~al.}, 2012, MNRAS, 422, 1835

\bibitem[{Shabala {et~al}\mbox{.}(2012)Shabala, Ting, Kaviraj, Lintott,
  Crockett, Silk, Sarzi, Schawinski, Bamford, \&
  Edmondson}]{2012MNRAS.423...59S}
Shabala S.~S. {et~al.}, 2012, MNRAS, 423, 59

\bibitem[{Shapiro {et~al}\mbox{.}(2010)Shapiro, Falc{\'o}n-Barroso, van~de Ven,
  de~Zeeuw, Sarzi, Bacon, Bolatto, Cappellari, Croton, Davies, Emsellem,
  Fakhouri, Krajnovi{\'c}, Kuntschner, McDermid, Peletier, van~den Bosch, \&
  van~der Wolk}]{2010MNRAS.402.2140S}
Shapiro K.~L. {et~al.}, 2010, MNRAS, 402, 2140

\bibitem[{Smith {et~al}\mbox{.}(2011)Smith, Dunne, Maddox, Eales, Bonfield,
  Jarvis, Sutherland, Fleuren, Rigby, Thompson, Baldry, Bamford, Buttiglione,
  Cava, Clements, Cooray, Croom, Dariush, De~Zotti, Driver, Dunlop, Fritz,
  Hill, Hopkins, Hopwood, Ibar, Ivison, Jones, Kelvin, Leeuw, Liske, Loveday,
  Madore, Norberg, Panuzzo, Pascale, Pohlen, Popescu, Prescott, Robotham,
  Rodighiero, Scott, Seibert, Sharp, Temi, Tuffs, van~der Werf, \& van
  Kampen}]{2011MNRAS.416..857S}
Smith D. J.~B. {et~al.}, 2011, MNRAS, 416, 857

\bibitem[{Smith {et~al}\mbox{.}(2012)Smith, Gomez, Eales, Ciesla, Boselli,
  Cortese, Bendo, Baes, Bianchi, Clemens, Clements, Cooray, Davies, De~Looze,
  di~Serego~Alighieri, Fritz, Gavazzi, Gear, Madden, Mentuch, Panuzzo, Pohlen,
  Spinoglio, Verstappen, Vlahakis, Wilson, \& Xilouris}]{2012ApJ...748..123S}
Smith M. W.~L. {et~al.}, 2012, ApJ, 748, 123

\bibitem[{Solomon {et~al}\mbox{.}(1997)Solomon, Downes, Radford, \&
  Barrett}]{1997ApJ...478..144S}
Solomon P.~M., Downes D., Radford S. J.~E., Barrett J.~W., 1997, ApJ, 478, 144

\bibitem[{Temi, Brighenti \& Mathews(2009)Temi, Brighenti, \&
  Mathews}]{2009ApJ...695....1T}
Temi P., Brighenti F., Mathews W.~G., 2009, ApJ, 695, 1

\bibitem[{Tremonti {et~al}\mbox{.}(2004)Tremonti, Heckman, Kauffmann,
  Brinchmann, Charlot, White, Seibert, Peng, Schlegel, Uomoto, Fukugita, \&
  Brinkmann}]{2004ApJ...613..898T}
Tremonti C.~A. {et~al.}, 2004, ApJ, 613, 898

\bibitem[{van Dokkum(2005)}]{2005AJ....130.2647V}
van Dokkum P.~G., 2005, AJ, 130, 2647

\bibitem[{Wei {et~al}\mbox{.}(2010)Wei, Vogel, Kannappan, Baker, Stark, \&
  Laine}]{Wei:2010bt}
Wei L.~H., Vogel S.~N., Kannappan S.~J., Baker A.~J., Stark D.~V., Laine S.,
  2010, ApJ, 725, L62

\bibitem[{Welch, Sage \& Young(2010)Welch, Sage, \& Young}]{Welch:2010in}
Welch G.~A., Sage L.~J., Young L.~M., 2010, ApJ, 725, 100

\bibitem[{Wilson(1995)}]{1995ApJ...448L..97W}
Wilson C.~D., 1995, ApJL, 448, L97

\bibitem[{Wright {et~al}\mbox{.}(2010)Wright, Eisenhardt, Mainzer, Ressler,
  Cutri, Jarrett, Kirkpatrick, Padgett, McMillan, Skrutskie, Stanford, Cohen,
  Walker, Mather, Leisawitz, Gautier, McLean, Benford, Lonsdale, Blain, Mendez,
  Irace, Duval, Liu, Royer, Heinrichsen, Howard, Shannon, Kendall, Walsh,
  Larsen, Cardon, Schick, Schwalm, Abid, Fabinsky, Naes, \&
  Tsai}]{2010AJ....140.1868W}
Wright E.~L. {et~al.}, 2010, AJ, 140, 1868

\bibitem[{Yi {et~al}\mbox{.}(2005)Yi, Yoon, Kaviraj, Deharveng, Rich, Salim,
  Boselli, Lee, Ree, Sohn, Rey, Lee, Rhee, Bianchi, Byun, Donas, Friedman,
  Heckman, Jelinsky, Madore, Malina, Martin, Milliard, Morrissey, Neff,
  Schiminovich, Siegmund, Small, Szalay, Jee, Kim, Barlow, Forster, Welsh, \&
  Wyder}]{2005ApJ...619L.111Y}
Yi S.~K. {et~al.}, 2005, ApJ, 619, L111

\bibitem[{Young \& Scoville(1991)}]{1991ARA&A..29..581Y}
Young J.~S., Scoville N.~Z., 1991, ARAA, 29, 581

\bibitem[{Young {et~al}\mbox{.}(2011)Young, Bureau, Davis, Combes, McDermid,
  Alatalo, Blitz, Bois, Bournaud, Cappellari, Davies, de~Zeeuw, Emsellem,
  Khochfar, Krajnovi{\'c}, Kuntschner, Lablanche, Morganti, Naab, Oosterloo,
  Sarzi, Scott, Serra, \& Weijmans}]{2011MNRAS.414..940Y}
Young L.~M. {et~al.}, 2011, MNRAS, 688

\bibitem[{Young {et~al}\mbox{.}(2014)Young, Scott, Serra, Alatalo, Bayet,
  Blitz, Bois, Bournaud, Bureau, Crocker, Cappellari, Davies, Davis, de~Zeeuw,
  Duc, Emsellem, Khochfar, Krajnovic, Kuntschner, McDermid, Morganti, Naab,
  Oosterloo, Sarzi, \& Weijmans}]{2014MNRAS.444.3408Y}
Young L.~M. {et~al.}, 2014, MNRAS, 444, 3408

\end{thebibliography}
\bibdata{1DETGbib.bib}
\bibstyle{mn2e}

\label{lastpage}
\end{document}